\documentclass[aap,seceqn,nameyear,MSNbibl,dvips]{arximspdf}

\doi{10.1214/09-AAP642}
\volume{20}
\issue{4}
\pubyear{2010}
\firstpage{1179}
\lastpage{1204}

\makeatletter

\newproclaim{Remark}{Remark}
\newproclaim{Assumption}{Assumption}

\newtheorem{Proposition}{Proposition}
\newtheorem{Theorem}{Theorem}

\makeatother

\begin{document}
\begin{frontmatter}

\title{On optimal arbitrage}
\runtitle{On optimal arbitrage}

\begin{aug}
\author[A]{\fnms{Daniel} \snm{Fernholz}\ead[label=e1]{fernholz@cs.utexas.edu}} and
\author[B]{\fnms{Ioannis} \snm{Karatzas}\corref{}\thanksref{T1}\ead[label=e2]{ik@enhanced.com}}
\runauthor{D. Fernholz and I. Karatzas}
\affiliation{University of Texas at Austin and INTECH}
\address[A]{Department of Computer Sciences\\
University of Texas at Austin\\
Austin, Texas 78712\\
USA\\
\printead{e1}}
\address[B]{INTECH Investment Management\\
One Palmer Square, Suite 441\\
Princeton, New Jersey 08542\\
USA\\
\printead{e2}}
\end{aug}

\thankstext{T1}{Author is on leave from the Department of Mathematics at Columbia University,
where his research was supported by NSF Grants DMS-06-01774 and DMS-09-05754.}

\received{\smonth{6} \syear{2008}}
\revised{\smonth{6} \syear{2009}}

%
\begin{abstract}
In a Markovian model for a financial market, we characterize the best
arbitrage with respect to the market portfolio that can be achieved
using nonanticipative investment strategies, in terms of the smallest
positive solution to a parabolic partial differential inequality; this
is determined entirely on the basis of the covariance structure of the
model. The solution is intimately related to properties of strict local
martingales and is used to generate the investment strategy which
realizes the best possible arbitrage. Some extensions to non-Markovian
situations are also presented.
\end{abstract}

%
\begin{keyword}[class=AMS]
\kwd[Primary ]{60H10}
\kwd{91B28}
\kwd[; secondary ]{60G44}
\kwd{35B50}.
\end{keyword}
\begin{keyword}
\kwd{Portfolios}
\kwd{arbitrage}
\kwd{parabolic operators}
\kwd{maximum principle}
\kwd{strict local martingales}
\kwd{exit measures for supermartingales}
\kwd{diffusions}
\kwd{Fichera drift}.
\end{keyword}

\end{frontmatter}

\section{Introduction}\label{sec1}

In a Markovian model for an equity market with mean rates of
return $ \mathrm{b}_{i } (\mathfrak{X}(t)) $ and covariance
rates $ \mathrm{a}_{i j} (\mathfrak{X}(t)) $, $1 \le i,j\le
n $, for its asset capitalizations $ \mathfrak{X}(t) = (
X_1(t),\ldots, X_n (t) )' \in(0,\infty)^n $ at time $ t $,
what is the highest return on investment [as in
(\ref{MaxRetInv}) below] that can be achieved relative
to the market on a given time--horizon $ [0,T] $, using
nonanticipative investment strategies? What are the weights
assigned to the different assets by such an investment strategy
that accomplishes this?

Answers: under suitable conditions, $ 1 / U (T,\mathfrak{X}(0)) $ and
\[
X_i (t) D_i \log U \bigl( T-t, \mathfrak{X}(t) \bigr) + { X_i
(t) \over X_1(t)+\cdots+ X_n (t) },\qquad i = 1,\ldots,
n, t \in[0,T],
\]
respectively. Here $ U\dvtx [0, \infty) \times(0, \infty)^n
\rightarrow(0,1] $ is the smallest nonnegative solution of the
linear parabolic partial differential inequality
%
%
\begin{equation}
\label{E.3}
{ \partial U \over\partial\tau} (\tau, \mathbf{x})
\ge\widehat{\mathcal{L}} U (\tau, \mathbf{x}),\qquad
(\tau, \mathbf{x}) \in(0, \infty) \times(0, \infty)^n,
\end{equation}
subject to the initial condition $ U (0, \cdot) \equiv1 $,
for the linear
operator
%
%
\begin{equation}
\label{E.3.a}
\widehat{\mathcal{L}}f := {1 \over2 } \sum_{i=1}^n
\sum_{j=1}^n x_i x_j \mathrm{a}_{ij} (\mathbf{x}) D^2_{ij}f +
\sum_{i=1}^n x_i \Biggl( \sum_{j=1}^n { x_j \mathrm{a}_{i j}
(\mathbf{x}) \over x_1 + \cdots+ x_n } \Biggr) D_{i}f
\end{equation}
with\vspace*{1pt} $ D_i = \partial
/\partial x_i $, $ D = ( \partial/\partial x_1,
\ldots,
\partial/\partial x_n )' $ and $ D^2_{ij} =
\partial^{ 2} / \partial x_i\, \partial x_j $.
Furthermore, $ U (T, \mathfrak{X} (0)) $ is the probability
that the $ ( [0,\infty)^n \setminus\{
\mathbf{0}\} )$-valued diffusion process $
\mathfrak{Y}(\cdot)= ( Y_1(\cdot), \ldots, Y_n (\cdot) )' $
with infinitesimal generator
$ \widehat{\mathcal{L}} $ as above and $ \mathfrak{Y} (0)
=\mathfrak
{X} (0)
\in(0, \infty)^n $ does not hit the boundary of the
orthant $ [0,\infty)^n $ by time $ t = T $.
We note that the answers
involve only the covariance structure of the market, not the
actual rates of return; the only role these latter play is to
ensure that the diffusion $ \mathfrak{X} (\cdot) $ lives in
$ (0, \infty)^n $.

Arbitrage relative to the market exists on $ [0,T] $, iff
$ U (T,\mathfrak{X}(0))<1 $; this
is deeply related to the importance of
\textit{strict local martingales} in the present context, and amounts to
failure of uniqueness for the Cauchy problem
\[
{\partial U \over\partial\tau} (\tau, \mathbf{x})
= \widehat{\mathcal{L}} U (\tau, \mathbf{x}),\qquad (\tau,
\mathbf{x}) \in(0, \infty) \times(0, \infty)^n \quad\mbox{and}\quad
U (0, \cdot) \equiv1 .
\]
Sufficient conditions for such failure of uniqueness
are provided.

Consider an ``auxiliary market'' with capitalizations $
\mathfrak{Y}(\cdot)= ( Y_1(\cdot),\ldots,\break Y_n (\cdot) )' $ as above. The
probabilistic significance of the change of drift inherent in the
definition of the operator $ \widehat{\mathcal{L}} $, from $
\mathrm{b}_i (\mathbf{x}) $ for $ \mathfrak{X}(\cdot) $ to $
\sum_{j=1}^n ( x_j \mathrm{a}_{i j} (\mathbf{x})) /\break ( x_1 + \cdots+ x_n
) $ for $ \mathfrak{Y}(\cdot) $, is that it corresponds to a change of
probability measure which makes the weights $ \nu_i (\cdot) :=Y_i
(\cdot)/( Y_1(\cdot)+\cdots+ Y_n (\cdot)), i = 1,\ldots, n $, of the
auxiliary market portfolio martingales. Its financial significance is
that it bestows to the auxiliary market portfolio $ \nu(\cdot) = (
\nu_1(\cdot),\ldots, \nu_n(\cdot) )' $ the so-called
\textit{num\'eraire property}: any strategy's relative performance in
the market with capitalizations $ \mathfrak{Y}(\cdot) $ is a
supermartingale, so this market cannot be outperformed. This change
need not come from a Girsanov-type (absolutely continuous)
transformation; rather it corresponds to, and represents, the
\textit{exit measure} of F\"ollmer (\citeyear{F72}) for an appropriate
supermartingale.

Sections \ref{sec2} and \ref{sec3} set up the model, whereas Section
\ref{sec4} introduces the notion and offers examples of relative
arbitrage; Section \ref{sec5} makes the connection with strict local
martingales. Section \ref{sec6} formulates the problem, and Section
\ref{sec7} offers some preliminary results, actually in some modest
generality (including non-Markovian cases). Section \ref{sec8} sets up
the Markovian model; the results are presented in earnest in Sections
\ref{sec9}--\ref{sec11}, Section \ref{sec12} discusses a couple of
examples in detail and a few open questions are raised in Section
\ref{sec13}.

\textit{Related literature}: the questions raised in this study are
related to the work of Delbaen and Schachermayer (\citeyear{DS1995b}).
They bear an even closer connection with issues raised in the Finance
literature under the general rubric of ``bubbles'' [see Definition 5
and Theorem 1 in Ruf (\citeyear{R09}) for the precise connection]. The
literature on this topic is large, so let us mention the papers by
Loewenstein and Willard (\citeyear{LW00}), Pal and Protter
(\citeyear{PP07}) and, most significantly, Heston, Loewenstein and
Willard (\citeyear{HLW07}), as the closest in spirit to our approach
here. We note the recent preprint by Hugonnier (\citeyear{H07}), which
demonstrates that arbitrage opportunities can arise in equilibrium
models; this preprint, and Heston, Loewenstein and Willard
(\citeyear{HLW07}), can be consulted for an up-to-date survey of the
literature on this subject and for some explicit computations of
trading strategies that lead to arbitrage. The need to consider
state-price-density processes that are only local (as opposed to true)
martingales has also been noticed in the context of ``stochastic
volatility'' models [e.g., Sin (\citeyear{S98}), Wong and Heyde
(\citeyear{WH06})] and of pricing with long maturities [e.g., Hulley
and Platen (\citeyear{HP08})].


\section{The model}\label{sec2}

We consider a model consisting of a money-market $
d B(t)=B(t) r(t)\, d t $, $ B(0)=1 $ and of $n$
stocks with capitalizations,
%
%
\begin{equation}
\label{1.1}\quad
d X_i(t) = X_i(t) \Biggl( \beta_i (t)\, d t +
\sum_{k=1}^K \sigma_{i k}(t) \,d W_k(t) \Biggr),\qquad X_i(0)=x_i>0,
\end{equation}
for $ i=1,\ldots, n $.
These are defined on a probability space $ (\Omega, \mathcal{F},
\mathbb{P}) $ and are driven by the Brownian motion $ W(\cdot) = (
W_1(\cdot),\ldots, W_K(\cdot) )' $ whose $K\ge n$ independent
components are the model's ``factors.''

We shall assume throughout that
the interest rate process of the money-market is $ r (\cdot)
\equiv0 $, identically equal to zero; and
that the vector-valued process $
\mathfrak{X}(\cdot) = ( X_1(\cdot), \ldots, X_n(\cdot)
)'$ of capitalizations, the vector-valued process $
\beta(\cdot) = ( \beta_1(\cdot), \ldots, \beta_n(\cdot)
)'$ of \textit{mean rates of return} for the various stocks and the
$(n \times K)$-matrix-valued process $\sigma(\cdot) = ( \sigma_{i
k}(\cdot) )_{1 \le i \le n, 1 \le k \le K } $ of
\textit{volatilities} are all progressively measurable with respect
to a right-continuous filtration $ \mathbb{F} = \{ \mathcal{F}(t)
\}_{0 \le t < \infty} $ which represents the ``flow of
information'' in the market with $ \mathcal{F}(0) = \{ \varnothing,
\Omega\} $, $\operatorname{mod}\mathbb{P} $. Let $ \alpha(\cdot) :=
\sigma(\cdot) \sigma^{\prime}(\cdot) $ be the \textit{covariance
process} of the stocks in the market, and impose for $ \mathbb{P}$-a.e.
$\omega\in\Omega$ the condition
%
%
\begin{equation}
\label{1.2}
\sum_{i=1}^n \int_0^T \bigl(
|\beta_i (t, \omega) | + \alpha_{ii} (t, \omega) \bigr)\,
d t < \infty\qquad \forall\ T \in( 0,\infty) .
\end{equation}
Under this condition the processes $ X_1 (\cdot),\ldots,
X_n ( \cdot) $ can be expressed as $ X_i ( \cdot)= x_i \exp
\{ \int_0^ \cdot( \beta_i (t) - {1 \over2 } \alpha_{ii}
(t) ) \,d t + \sum_{k=1}^K \int_0^ \cdot\sigma_{ik}(t) \,dW_k (t)
\} > 0 $.

In this setting, the Brownian motion $ W (\cdot) $ need not be adapted
to the ``observations'' filtration $ \mathbb{F} $. It \textit{is}
adapted, though, to the $ \mathbb{P}$-augmentation $ \mathbb{G} = \{
\mathcal{G} (t) \}_{0 \le t < \infty} $ of the filtration $ \mathbb{F}
$, provided that $ K=n $ and that the matrix-valued process
$\sigma(\cdot) $ is invertible---as in Assumption \ref{AssumptionB}
below.


\section{Strategies and portfolios}\label{sec3}

Consider now a \textit{small investor} who decides, at each time $ t $,
which proportion $ \pi_i (t) $ of current wealth $ V(t)$ to
invest in
the $i$th stock, $ i = 1,\ldots, n $; the proportion $ 1 - \sum
_{i=1}^n \pi_i
(t) =: \pi_0 (t) $ gets invested in the money market. Thus, the
wealth $ V(\cdot) \equiv V^{v, \pi} (\cdot) $ for an initial
capital $ v \in(0, \infty) $ and an investment strategy $
\pi(\cdot) = ( \pi_1 (\cdot),\ldots, \pi_n (\cdot)
)' $ satisfies the initial condition $ V (0) =v $ and
%
%
\begin{eqnarray}
\label{1.8}
\frac{ d V (t)}{V (t)} &=&
\sum_{i=1}^n \pi_i(t) \,\frac{ d X_i(t)}{X_i(t)}
+ \pi_0(t) \,\frac{ d B(t)}{B(t)} \nonumber\\[-8pt]\\[-8pt]
&=& \pi'(t) [ \beta(t)\, d t +
\sigma(t)\, d W (t) ] .\nonumber
\end{eqnarray}
We shall call \textit{investment strategy} a $
\mathbb{G}$-progressively measurable process $ \pi\dvtx [0$,\break
$\infty) \times\Omega\rightarrow\mathbb{R}^n $ which satisfies for $
\mathbb{P}$-a.e. $ \omega\in\Omega$ the analogue
\[
 \int_0^T \bigl( | \pi'
(t, \omega) \beta(t,\omega) | + \pi' (t, \omega)
\alpha(t,\omega) \pi(t, \omega) \bigr)\, d t < \infty,\qquad  \forall\, T \in(0,
\infty)
\]
of (\ref{1.2}). The collection of investment strategies will
be denoted by $ \mathcal{H} $.

A strategy $ \pi(\cdot) \in\mathcal{H} $ with $ \sum_{i=1}^n \pi_i (t,
\omega) = 1 $ for all $ (t, \omega) \in[0,\infty) \times\Omega $ will
be called \textit{portfolio}. A portfolio never invests in the money
market and never borrows from it. We shall say that a process $
\pi(\cdot) $ is \textit{bounded}, if for it there exists a real
constant $ C_\pi>0 $ such that $ \| \pi(t,\omega)\| \le C_\pi$ holds
for all $ (t, \omega) \in [0,\infty) \times\Omega$. We shall call
\textit{long-only portfolio} one that satisfies $ \pi_1 (t, \omega)
\ge0,\ldots, \pi_n (t, \omega) \ge0 , \forall(t, \omega) \in[0,\infty)
\times\Omega $, that is, never sells any stock short. Clearly, a
long-only portfolio is also bounded.

Corresponding to an investment strategy $ \pi(\cdot) $ and
initial capital $ v >0 $, the associated wealth process,
that is, the solution of (\ref{1.8}), is
\[
V^{v, \pi} (\cdot)= v \exp\biggl\{ \int_0^\cdot\pi' (t) \biggl(
\beta
(t)- { \alpha(t) \over2 } \pi(t) \biggr)\,d t + \int_0^\cdot\pi'
(t) \sigma(t) \,d W(t) \biggr\} > 0 .
\]
The strategy $ \varrho(\cdot) \equiv0 $ invests only in the
money market at all times; it results in $ V^{v, \varrho} (
\cdot) \equiv v $, that is, in hoarding the initial wealth under
the mattress.

\subsection{The market portfolio}\label{sec31}

An important long-only portfolio is the market
portfolio; this invests in all stocks in proportion
to their relative weights,
%
%
\begin{equation}
\label{2.1}\qquad
\mu_i(t) := \frac{X_i(t)}{X(t)},\qquad i =1,\ldots, n,
\mbox{where } X(t) := X_1(t) + \cdots+ X_n(t) .
\end{equation}
Clearly $ V^{v, \mu}(\cdot)=v X (\cdot) / X(0) $, and the resulting
vector process $ \mu(\cdot) = ( \mu_1 (\cdot)$,\break $\ldots,\mu_n
(\cdot) )' $ of market weights takes values in the positive simplex $
\Delta^n_+ := \{ ( m_1,\ldots, m_n)' \in(0,1)^n | \sum_{i=1}^n m_i  = 1
\} $ of $ \mathbb{R}^n $. An application of It\^o's rule gives, after
some computation, the dynamics of this process as
%
%
\begin{equation}
\label{2.2}
d \mu_i (t) = \mu_i (t) \Biggl[ \gamma_i^\mu(t) \,d t +
\sum_{k=1}^K \tau^\mu_{i k} (t) \,d W_k (t) \Biggr],\qquad
i = 1,\ldots, n .
\end{equation}
Here $ \tau^\mu(t) $ is the matrix with entries $ \tau^\mu_{i k} (t) :=
\sigma_{i k } (t) - \sum_{j=1}^n \mu_j (t) \sigma_{j k } (t) $, $
\mathfrak{e}_i $ the $i$th unit vector in $ \mathbb{ R}^n$ and the
vector $ \gamma^\mu(t) := ( \gamma_1^\mu(t),\ldots, \gamma_n^\mu(t) )'
$ has
%
%
\begin{equation}
\label{2.4}
\gamma_i^\mu(t) := \bigl( \mathfrak{e}_i - \mu(t) \bigr)' \bigl(
\beta(t) - \alpha(t) \mu(t) \bigr) .
\end{equation}


\section{Relative arbitrage}\label{sec4}

The following notion was introduced in Fernholz (\citeyear{F02}): given
a real number $ T>0 $ and any two investment strategies $ \pi(\cdot) $
and $ \rho(\cdot) $, we call $\pi(\cdot)$ an \textit{arbitrage relative
to} $\rho(\cdot)$ over $[0,T]$, if
%
%
\begin{equation}
\label{AO}
{\mathbb P} \bigl( V^{1,\pi}(T) \ge V^{1,\rho} (T)
\bigr) =1 \quad\mbox{and}\quad {\mathbb P} \bigl( V^{1,\pi}(T) >
V^{1,\rho}(T) \bigr)>0 .
\end{equation}
We call such relative arbitrage \textit{strong} if
$ {\mathbb P} ( V^{1,\pi}(T) >
V^{1,\rho}(T) ) =1 $.

Arbitrage (resp., strong arbitrage) relative to $ \varrho
(\cdot) \equiv0 $ that invests only in the money market, is
called just that, without the qualifier ``relative.''

\subsection{Examples of arbitrage relative to the market}\label{sec41}

Here are some examples taken from the survey Fernholz and Karatzas
(\citeyear{FK09}), especially Sections 7 and 8, Remark 11.4, Examples
11.1 and 11.2. Suppose first that
%
%
\begin{equation}
\label{IntVol} \sum_{i=1}^n \mu_i (t) \alpha_{ii}(t) - \sum_{i=1}^n
\sum_{j=1}^n \mu _i (t) \alpha_{ij}(t)\mu_j (t) \ge h \qquad \forall\,0
\le t < \infty
\end{equation}
holds almost surely for some constant $ h >0 $. Then the long-only
portfolio $ \pi_i (t) = \mu_i (t) ( c - \log\mu_i (t) )/J (t) , i=
1,\ldots, n $, $ J(t):=\sum_{j=1}^n \mu_j (t) ( c - \log\mu_j (t)) $
is, for sufficiently large $ c >0 $, a strong arbitrage relative to the
market portfolio $ \mu (\cdot) $ over any time--horizon $[0,T]$ with $ T
> (2 \log n )/ h $.

Another condition guaranteeing the existence of strong arbitrage
relative to the market is that there exists a real constant $ h
>0 $ with
%
%
\begin{equation}\label{EW}\qquad
\sqrt[n]{\mu_1 (t) \cdots\mu_n (t) } \Biggl[ \sum_{i=1}^n \alpha_{ii} (t)- {
1 \over n } \sum_{i=1}^n \sum_{j=1}^n \alpha_{ij} (t) \Biggr] \ge h
\qquad \forall\,0 \le t < \infty
\end{equation}
a.s. Then for $ c
>0 $ sufficiently large, the long-only portfolio
$ \pi_i (t)= \lambda(t) (1/n) + (1 - \lambda(t)) \mu_i
(t) $, $1 \le i \le n $, $ 1 / \lambda(t) := 1+ ( ( \mu_1 (t)
\cdots\mu_n (t) )^{1/n}/c ) $,
is strong arbitrage relative to the market over any
$[0,T]$ with $ T > (2 n^{1 - (1/n)} )/ h $.
\begin{Remark}\label{Remark1}
Suppose that all the eigenvalues of the covariance matrix-valued
process $ \alpha(\cdot) $ are bounded away from both zero and infinity,
uniformly on $[0, \infty) \times\Omega$, and that (\ref {IntVol})
holds. Then, for any given constant $ p \in(0,1) $, the long-only
portfolio $ \mu_i^{(p)} (t) = (\mu_i (t) )^p ( \sum_{j=1}^n (\mu_j (t)
)^p )^{-1} $, $ i= 1,\ldots, n $, leads again to strong arbitrage
relative to the market portfolio over sufficiently long time--horizons.
It is also of great interest that appropriate modifications of the
portfolio $ \mu^{(p)} (\cdot) $ yield such arbitrage over \textit{any}
time--horizon $[0,T]$.
\end{Remark}


\section{Market price of risk and strict local martingales}\label{sec5}

We shall assume from now on that there exists a \textit{market price
of risk} $ \vartheta\dvtx [0,\infty) \times\Omega\rightarrow
\mathbb{R}^K $, an $ \mathbb{F}$-progres\-sively measurable
process that satisfies
%
%
\begin{eqnarray}
\label{TH}
&\sigma(t, \omega) \vartheta(t, \omega) = \beta(t, \omega) \qquad
\forall( t, \omega) \in[0,\infty) \times\Omega
\quad\mbox{and}&
\nonumber\\[-8pt]\\[-8pt]
&\displaystyle\mathbb{P} \biggl( \int_0^T
\|\vartheta(t, \omega) \|^2 \,d t < \infty,\,\forall\, T
\in(0,\infty) \biggr) = 1.&\nonumber
\end{eqnarray}
The existence of a
market-price-of-risk process $ \vartheta(\cdot) $ allows us to
introduce an associated exponential local martingale,
%
%
\begin{equation}
\label{Z}\qquad
Z(t) := \exp\biggl\{ - \int_0^t
\vartheta' (s) \,d W(s) - { 1 \over2} \int_0^t
\| \vartheta(s) \|^2 \,d s \biggr\},\qquad 0 \le t < \infty.
\end{equation}
This process is also a supermartingale; it is a martingale, if and only
if $ \mathbb{E} (Z(T))=1 $ holds for all $ T \in(0,\infty)$. For the
purposes of this work it is important to allow such exponential
processes to be strict local martingales; that is, not to exclude the
possibility $ \mathbb{E} (Z(T))<1 $ for some $ T \in(0,\infty)$.

From (\ref{Z}) and (\ref{1.8}), now written in the form
%
%
\begin{equation}
\label{V}\qquad\quad
d V^{v, \pi} ( t) = V^{v, \pi} ( t) \pi'(t)\sigma(t) \,
d\widehat{W} (t) ,\qquad \widehat{W} (t) := W (t)+
\int_0^t \vartheta(s) \,d s
\end{equation}
on the strength of (\ref{TH}), the product rule of It\^o's calculus
shows that
%
%
\begin{equation}
\label{1.10}
Z(\cdot) V^{v, \pi} ( \cdot)= v + \int_0^\cdot Z(t) V^{v, \pi} (t)
\bigl( \sigma' (t) \pi(t) - \vartheta(t) \bigr)' \,d W (t)
\end{equation}
is a positive local martingale
and a supermartingale, for every $ \pi(\cdot) \in
\mathcal{H} $.

If $ \alpha(\cdot)$ is invertible, we can take $ \vartheta(\cdot)=
\sigma^{\prime} (\cdot) \alpha^{-1} (\cdot) \beta(\cdot) $ as market
price of risk in (\ref{TH}). If $ \beta(\cdot) = \alpha(\cdot)
\mu(\cdot) $ holds we can select $ \vartheta(\cdot) = \sigma'
(\cdot)\mu (\cdot) $ and get $ Z(\cdot) \equiv v / V^{v, \mu} (\cdot)
\equiv X(0) / X(\cdot) $ from (\ref{1.10}); there is then no arbitrage
relative to the market because $ V^{v, \pi} (\cdot) / V^{v, \mu}
(\cdot) $ \textit{is a supermartingale for all} $ \pi(\cdot)
\in\mathcal{H } $; thus $ \mathbb{E} [ V^{1,\pi}(T) / V^{1,\mu}(T)
]\le1 $, a conclusion at odds with (\ref{AO}).


\subsection{Strict local martingales}\label{sec51}

Suppose the covariance process $ \alpha(\cdot) $ is\break bounded,
and (\ref{AO}) holds for two bounded portfolios $ \pi
(\cdot) $ and $ \rho(\cdot) $. Then, for \textit{any}
market-price-of-risk process $ \vartheta(\cdot) $ as in
(\ref{TH}), the positive local martingales $ Z(\cdot) $ and
$ Z(\cdot) V^{v, \rho}(\cdot) $ of (\ref{Z}), (\ref{1.10}) are
strict: $ \mathbb{E} [
Z(T) V^{v, \rho}(T) ] < v $, $ \mathbb{E} (Z(T) ) <1 $
[Fernholz and Karatzas (\citeyear{FK09}), Section 6].

In particular, if the matrix $ \alpha(\cdot) $ is bounded, and
(\ref{AO}) holds for some bounded portfolio $ \pi(\cdot) $ and
for the market portfolio $ \rho(\cdot) \equiv\mu(\cdot) $
(these assumptions are satisfied, e.g., under the conditions
in Remark \ref{Remark1}), then
%
%
\begin{eqnarray}
\label{AO.b}
&&\mathbb{E} (Z(T) )<1 ,\qquad
\mathbb{E} [Z(T) X(T) ]< X(0) ,
\nonumber\\[-8pt]\\[-8pt]
&&\mathbb{E} [ Z(T) X_i(T) ] < X_i(0),\qquad i=1,\ldots, n .\nonumber
\end{eqnarray}


\section{Optimal arbitrage relative to the market}\label{sec6}

The possibility of strong arbitrage relative to the market,
defined and exemplified in Section \ref{sec4}, raises an obvious question:
\textit{what is the best possible arbitrage of this kind}?

One
way to cast this question is as
follows: on a given time--horizon $ [0,T] $, what is the
\textit{smallest} relative amount,
%
%
\begin{equation}
\label{C.3}
\mathfrak{u} (T) := \inf\bigl\{ w >0 \,|\, \exists\,
\pi(\cdot) \in\mathcal{H} \mbox{ s.t. } V^{ w X(0),\pi} (T) \ge
X(T), \mbox{ a.s.} \bigr\},
\end{equation}
of initial capital, starting with which one can match or exceed at
time $ t = T $ the market capitalization $ X(T) $? Clearly,
$
0< \mathfrak{u} (T) \le1 $;
and for $ 0<w <\mathfrak{u} (T) $, no strategy starting with
initial capital $ w X(0) $ can outperform the market almost surely,
over the
horizon $ [0,T] $. That is, for every $ \pi(\cdot) \in
\mathcal{ H} $ and $ 0<w <\mathfrak{u} (T) $, we have $
\mathbb{P} [ V^{ w X(0), \pi} (T) \ge X(T) ] <
1 $.

We shall impose from now on the following structural assumptions on
the filtration $ \mathbb{F} = \{ \mathcal{F}(t) \}_{0 \le t <
\infty} $, the ``flow of information'' in the market.
\renewcommand{\theAssumption}{A}
\begin{Assumption}\label{AssumptionA}
Every local martingale of the
filtration $ \mathbb{F}$ can be represented as a stochastic
integral, with respect to the driving Brownian motion $ W
(\cdot) $ in (\ref{1.1}), of some $ \mathbb{G}$-progressively
measurable integrand.
\end{Assumption}
\renewcommand{\theAssumption}{B}
\begin{Assumption}\label{AssumptionB}
We have $ K=n $, and
$ \sigma(t) $ is invertible, $ \forall t
\in[0,T] $.
\end{Assumption}

Under these two assumptions, general results about hedging in
so-called complete markets [e.g., Karatzas and Shreve (\citeyear{KS1998}),
Fernholz and Karatzas (\citeyear{FK09}), Section 10 or Ruf (\citeyear{R09})] based on
martingale representation results, show that the quantity of
(\ref{C.3}) given as
%
%
\begin{equation}
\label{C.5}
\mathfrak{u} (T) =
\mathbb{E} [ Z(T) X(T) ] / X(0);  \qquad\mbox{that }
V^{ \mathfrak{u} (T) X(0), \widehat{\pi}} (T) = X(T)
\end{equation}
holds a.s. for some $ \widehat{\pi} (\cdot) \in\mathcal{H} $;
and that $ 1 / \mathfrak{u} (T) $ gives the highest return,
%
%
\begin{equation}
\label{MaxRetInv}
\sup\{ q \ge1 |\, \exists\,\pi(\cdot) \in\mathcal{H}
\mbox{ s.t. } V^{ 1,\pi} (T) \ge q V^{
1,\mu}(T)\mbox{, a.s.} \},
\end{equation}
on investment, that one can achieve relative to the market over
$ [0,T] $. Arbitrage relative to the market is possible on
$ [0,T] $, if and only if $ \mathfrak{u} (T) <1 $.

The result in (\ref{C.5}) provides no information about the
strategy $ \widehat{\pi} (\cdot) $
that implements this ``best possible'' arbitrage, apart from
ascertaining its existence. In Section \ref{sec8} we
shall specialize the model of (\ref{1.1}) to a Markovian context
and describe $ \widehat{\pi} (\cdot) $ in terms of partial
differential equations (Section \ref{sec11}). We shall also characterize
the quantity $ \mathfrak{u} (T) $ in terms of the
smallest solution to a parabolic partial differential inequality,
and as the probability of nonabsorption by time $T$ for a
suitable diffusion (Theorems \ref{Theorem1}, \ref{Theorem2}).

Assumption \ref{AssumptionA} holds when $ \mathbb{F} $
is (the augmentation of) $ \mathbb{F}^W $, the filtration
generated by the Brownian motion $ W (\cdot) $; as well as
when Assumption \ref{AssumptionB} holds, the
$ \beta_i (
\cdot) , \sigma_{i \nu} ( \cdot) $ are all progressively
measurable with respect to $ \mathbb{F}^\mathfrak{X} = \{ \mathcal
{F}^\mathfrak{X}(t) \}_{0 \le t < \infty} $, $ \mathcal
{F}^\mathfrak
{X}(t):= \sigma( \mathfrak{X} (s), 0 \le s \le
t) $, and $ \mathbb{F} \equiv\mathbb{F}^\mathfrak{X}_+ $
[Jacod (\citeyear{J77})].


\subsection{Generalized likelihood ratios}\label{sec61}

The positive local martingale $ Z(\cdot) X(\cdot)$, whose
expectation appears in (\ref{C.5}), can be expressed as
%
%
\begin{equation}
\label{C.7.aa}
Z( t) X( t) = X(0)\cdot\exp\biggl\{ - \int_0^t
( \widetilde{\vartheta} (s) )' \,d W(s) - { 1 \over2}
\int_0^t
\| \widetilde{\vartheta} (s) \|^2 \,d s \biggr\}
\end{equation}
for $ 0 \le t \le T $. Here we have solved equation (\ref{1.10})
for $ \pi(\cdot) \equiv\mu(\cdot) $ and set
%
%
\begin{equation}
\label{C.8}
\widetilde{\vartheta} (\cdot) := \vartheta(\cdot) -
\sigma' (\cdot) \mu(\cdot) ,\qquad \widetilde{W}(\cdot) :=
W(\cdot) + \int_0^\cdot\widetilde{\vartheta} (t ) \,d t ,
\end{equation}
whence $ \sigma(\cdot) \widetilde{\vartheta} (\cdot) = \beta
(\cdot) - \alpha(\cdot) \mu(\cdot)$ from (\ref{TH}); we thus re-cast
(\ref{1.1}) as
%
%
\begin{equation}
\label{C.9}
d X_i (t) = X_i (t) \Biggl[ { \sum_{j=1}^n \alpha_{i j}
(t) X_j (t) \over X_1 (t) + \cdots+ X_n (t) }\, dt +
\sum_{k=1}^n \sigma_{i k} (t) \,d \widetilde{W}_k (t) \Biggr] .
\end{equation}

On the other hand, we note from (\ref{C.7.aa}), (\ref{C.8}) that
the reciprocal of the exponential local martingale
$ Z(\cdot) X(\cdot) / X(0) $ can be expressed as
%
%
\begin{equation}
\label{C.7.bb}
\Lambda(\cdot) := { X (0) \over Z(\cdot) X (\cdot) }
=
\exp\biggl\{ \int_0^\cdot
( \widetilde{\vartheta} (t) )' \,d \widetilde{W}(t) - { 1
\over
2} \int_0^\cdot
\| \widetilde{\vartheta} (t) \|^2 \,d t \biggr\};
\end{equation}
similarly,
the reciprocal of the local martingale $ Z(\cdot) X_i(\cdot)
/ X_i (0) $ is
%
%
\begin{equation}
\label{C.7.bba}\qquad
\Lambda_i(\cdot) := { X_i(0) \over Z(\cdot) X_i(\cdot) } =
\exp\biggl\{ \int_0^\cdot
\bigl( \widetilde{\vartheta}^{(i)} (t) \bigr)'\,
d \widetilde{W}^{(i)}(t) - { 1 \over2} \int_0^\cdot
\bigl\| \widetilde{\vartheta}^{(i)} (t) \bigr\|^2 \,d t \biggr\},
\end{equation}
where $ \widetilde{\vartheta}^{(i)} (\cdot) := \vartheta(\cdot) -
\sigma' (\cdot) \mathfrak{e}_i $ and $ \widetilde{W}^{(i)} (\cdot) :=
W(\cdot) + \int_0^\cdot\widetilde{\vartheta}^{(i)} (t ) \,dt$.

Comparing (\ref{C.7.bb}) and (\ref{C.7.bba}), we observe that $ \mu_i
(0) \Lambda(\cdot) = \mu_i (\cdot) \Lambda_i (\cdot) $ and cast the
dynamics of (\ref{2.2}) and (\ref{2.4}) for the market portfolio $ \mu
(\cdot) $ as
%
%
\begin{equation}
\label{2.2.a}
d \mu_i (t) = \mu_i (t) \bigl( \mathfrak{e}_i - \mu(t) \bigr)'
\sigma
(t) \,d \widetilde{W} (t),\qquad i=1,\ldots, n.
\end{equation}

If $ \mathfrak{u} (T) =1 $, that is, $ Z(\cdot) X(\cdot) $ is a
martingale on $ [0,T]$, no arbitrage relative to the market is possible
on this time--horizon; the ``reference'' measure
%
%
\begin{equation}
\label{C.9.a}
\widetilde{\mathbb{P}}_T (A) :=
\mathbb{E} [ Z(T) X(T) 1_A ] / X(0) ,\qquad
A\in \mathcal{F} (T),
\end{equation}
is a probability, that is, $ \mathfrak{u}
(T)=\widetilde{\mathbb{P}}_T (\Omega)=1 $; and under
$ \widetilde{\mathbb{P}}_T $, the process $ \widetilde{W}(t),
0 \le t \le T $, in (\ref{C.8}) is a Brownian motion
by the Girsanov theorem, so from (\ref{2.2.a}) the market weights $
\mu_1
( t),\ldots, \mu_n ( t) , 0 \le t \le T $ are martingales.

We shall characterize next $ \mathfrak{u} (T) $ in terms
of the \textit{F\"ollmer exit measure}, of a ``generalized martingale
measure'' and of a measure
$ \mathbb{Q} $ with respect to which $ \mathbb{P} $
is locally absolutely continuous [equations (\ref{C.6a}),
(\ref{C.6b})] and which plays, to a considerable extent, the
r\^ole of $ \widetilde{\mathbb{P}}_T $ when $ Z(\cdot)
X(\cdot) $ fails to be a $ \mathbb{P}$-martingale. The
processes of (\ref{C.7.aa})--(\ref{C.7.bba}) are important in this
effort.


\section{Exit measure of a positive supermartingale}\label{sec7}

We shall assume in this section that the process $
Z(\cdot) $ of (\ref{Z}) is adapted to
$ \mathbb{F} = \{ \mathcal{F} (t)\}_{0 \le t <
\infty} $ and that this filtration is, in turn, the
right-continuous version $ \mathcal{F} (t) = \bigcap_{\varepsilon
>0} \mathcal{F}^{o} (t+\varepsilon) $ of a \textit{standard system}
$ \mathbb{F}^{o} = \{ \mathcal{F}^o (t)\}_{0 \le t < \infty}
$: to wit, each $ (\Omega, \mathcal{F}^o(t)) $ is isomorphic to the
Borel $\sigma$-algebra of some Polish
space, and for any decreasing sequence $ \{A_j\}_{j \in
\mathbb{N}} $ such that $A_j$ is an atom of $
\mathcal{F}^o(t_j)$, for some increasing sequence $ \{t_j\}_{j
\in\mathbb{N}} \subset[0,\infty) $, we have $ \bigcap_{j \in
\mathbb{N}} A_j \neq\varnothing$.

The canonical example is the space $ \Omega$ of right-continuous
paths $ \omega\dvtx [0,\infty) \rightarrow\mathbb{R}^n \cup\{ \Delta
\}
$, where $ \Delta$ is an additional ``absorbing point''; paths
stay at $ \Delta$ once they get there, that is, after $ \mathcal
{T} (\omega) = \inf\{ t \ge0 | \omega(t) = \Delta\} $, and are
continuous on $ (0, \mathcal{T} (\omega))$. If $ \mathcal{F}^o (t)
=\sigma(\omega(s), 0\le s \le t) $, then $ \mathbb{F}^{o} = \{
\mathcal{F}^o (t)\}_{0 \le t < \infty}
$ is a standard system [see F\"ollmer (\citeyear{F72}), the Appendix].

Under these conditions, we can associate to the $(\mathbb{P},
\mathbb{F})$-local martingale $ Z(\cdot)\cdot X(\cdot ) $ a
positive measure $ \mathfrak{P} $ on the predictable $ \sigma$-algebra
of $ [0, \infty] \times\Omega$,
\[
\mathfrak{P} \bigl( (T, \infty] \times A \bigr) :=
\mathbb{E} [ Z(T) X(T) 1_A ] / X(0) ,\qquad
A \in\mathcal{F} (T), T \in[0 ,\infty) ,
\]
by invoking an extension result [Parthasarathy (\citeyear{P67}),
Theorem V.4.1, whence the assumptions on the nature of the probability
space].

This is the ``exit measure'' of the supermartingale $ Z(\cdot)
X(\cdot) $, introduced by F\"ollmer (\citeyear{F72}, \citeyear{F73}) [see also Delbaen and
Schachermayer (\citeyear{DS1995a}), F\"ollmer and Gundel (\citeyear{FG06})]. F\"ollmer (\citeyear{F72})
obtained a characterization of the (process-theoretic)
properties of supermartingales, such as $
Z(\cdot) X(\cdot) $ here, in terms of the properties of $
\mathfrak
{P} $. It follows from his work that
$ Z(\cdot) X(\cdot) $ is a:

$\bullet$ \textit{martingale}, if and only if
$ \mathfrak{P} $ in concentrated on $ \{\infty\} \times
\Omega$;

$\bullet$ \textit{potential} [i.e., $
\mathfrak{u}(\infty)=0$], if and only if $ \mathfrak{P} $ in
concentrated on $ (0, \infty) \times\Omega$.


\subsection{\texorpdfstring{A representation of the F\"ollmer measure}{A representation of the Follmer measure}}\label{sec71}

From Theorem 4 in Delbaen and Schachermayer
(\citeyear{DS1995a}) and Theorem 1 and Lemma 4 of Pal and Protter (\citeyear{PP07}), the
process $ \Lambda(\cdot) $ of (\ref{C.7.bb})
is a continuous martingale under
some probability measure $ \mathbb{Q} $ on the filtered
space $ (\Omega, \mathcal{F}), \mathbb{F} $ as above.
The measure $ \mathbb{P} $ is locally absolutely continuous
with respect to $ \mathbb{Q} $, with $ d \mathbb{P}= \Lambda
(T) \,d \mathbb{Q} $ on each $ \mathcal{F} (T)$; and the
process $ \widetilde{W}(\cdot) $ of (\ref{C.8}) is $
\mathbb{Q}$-Brownian motion [cf. Ruf (\citeyear{R09}), Section 5]. Thus, from
(\ref{2.2.a}) \textit{the weights $ \mu_1 (\cdot),\ldots, \mu
_n(\cdot)
$ are
martingales} and satisfy $ \sum_{i=1}^n \mu_i (\cdot) \equiv1 $
a.e., under $ \mathbb{Q} $.

We consider the first time the process $ \Lambda(\cdot) $ hits the origin,
%
%
\begin{equation}
\label{Ttau}
\mathcal{ T} :=
\inf\{ t \ge0 | \Lambda(t)=0\} = \inf\{ t \ge0 | Z (t)
X(t)=\infty\}
\end{equation}
(infinite, if the set is empty). We have $ \mathbb{P}( \mathcal{T}
<\infty)=0 $, but
$ \mathbb{Q} (\mathcal{T} <\infty) $ can be
positive, so $ \mathbb{Q} $ may not be absolutely continuous with
respect to $ \mathbb{P} $; whereas, $
\mathbb{Q} $-a.e. on $\{ \mathcal{T} <\infty\}$, we have $
Z(\mathcal{T}+h) X(\mathcal{T}+h) = \infty$, $\forall h \ge
0 $
and $ \int_0^\mathcal{T} \|\widetilde{\vartheta} (t) \|
^2 \,d
t=\infty$. Intuitively, the role of the absorbing state $ \Delta$
is to
account for events that have zero $ \mathbb{P}$-measure, but positive
$ \mathbb{Q}$-measure. We also introduce the first times the
processes $ \mu_i (\cdot) $ and $
\Lambda_i (\cdot) $ hit the origin,
%
%
\begin{equation}
\label{TtauToo}
\mathcal{ T}_i :=\inf\{ t \ge0 | \mu_i (t)=0\} ,\qquad
\mathcal{\widetilde{T}}_i :=
\inf\{ t \ge0 | \Lambda_i (t)=0\} .
\end{equation}
\begin{Proposition}\label{Proposition1}
\textup{(i)} The quantity of (\ref{C.3}) can be
represented as
%
%
\begin{equation}
\label{C.6a}
\mathfrak{u} (T) = \mathfrak{P} \bigl( (T, \infty] \times\Omega
\bigr) = \mathbb{Q} ( \mathcal{T} > T ) .
\end{equation}

\textup{(ii)} Suppose $n \ge2$ and that all capitalizations $ X_1 (\cdot
),\ldots, X_n (\cdot)$ are real-valued $ \mathbb{Q}$-a.e. Then we
also have the $ \mathbb{Q}$-a.e. representations
%
%
\begin{equation}
\label{C.6aa}\qquad
\mathcal{T}=\min_{1 \le i \le n} \widetilde{\mathcal{T}}_i;
\quad
\mbox{as well as}\quad \mathcal{T}=\min_{1 \le i \le n} \mathcal
{T}_i \qquad\mbox{away from the event } E ,
\end{equation}
where $ E:= \{ \mathcal{T} <\infty\} \cap\{ \mu_1( \mathcal{T}
) \cdots\mu_n( \mathcal{T})>0\} $. This event has $ \mathbb
{Q}$-measure equal to zero, if for some real constant $ C >0 $ we have
%
%
\begin{equation}
\label{C.6w}
\| \vartheta(t, \omega) \|^2 \le C \bigl( 1 + \operatorname{Tr} (
\alpha(t, \omega) ) \bigr) \qquad \forall(t, \omega) \in[0,
\infty) \times\Omega.
\end{equation}
\end{Proposition}
\begin{pf}
We note $ \mathfrak{P} ( (T, \infty] \times A ) =
\mathbb
{E}^\mathbb{P} ( \Lambda^{-1} (T) 1_{
A \cap\{ \mathcal{T} > T \} } ) = \mathbb{E}^\mathbb{Q} (
\Lambda(T) \cdot\Lambda^{-1} (T) 1_{ A \cap\{ \mathcal{T} > T \}
} )= \mathbb{Q} (A \cap\{ \mathcal{T} > T \} ) $, $
\forall A \in\mathcal{F}(T)$. With $ A = \Omega$, we get
(\ref
{C.6a}). For $ A = \{ \mu_1 (T) \cdots\mu_n (T) =0\} $, this gives
$ \mathbb{Q} (A \cap\{ \mathcal{T} > T \} )=0 $: all the
weights $ \mu_1 (\cdot),\ldots, \mu_n (\cdot)$ are strictly positive
[equivalently, all $ X_1 (\cdot),\ldots, X_n (\cdot)$ take values in
$(0, \infty)$] on $[0, \mathcal{T})$, $ \mathbb{Q}$-a.e.

Recall $ \mu_i (0) \Lambda(\cdot) \equiv\mu_i (\cdot) \Lambda_i
(\cdot) $, $\forall\, i=1,\ldots, n $ from (\ref{C.7.bb}), (\ref
{C.7.bba}); this gives\break $ 1 / \Lambda(\cdot) =\sum_{i=1}^n (
\mu
_i (0) / \Lambda_i (\cdot)) $ on $ [0, \mathcal{T})$, and the first
equation in (\ref{C.6aa}).

On the event $ \{ \mathcal{T} < \infty\} \setminus E $, for some $j
\in\{ 1,\ldots, n\}$ we shall have $\mu_j (\mathcal{T}) =0$, thus
also $ \mathcal{T}_j = \mathcal{T}$ and
the second equation in (\ref{C.6aa}). On the other hand, we have seen
that $ \mathcal{T}_i=\infty$, $\forall i=1,\ldots, n $
holds $ \mathbb{Q}$-a.e. on $ \{ \mathcal{T} = \infty\} $, so this
equation is valid on $\{ \mathcal{T} =\infty\}$.

Finally, from (\ref{C.9}), (\ref{C.7.bb}): $ \int_0^\mathcal{T}
\operatorname{Tr} ( \alpha(t, \omega) ) \,dt < \infty$, $
\int
_0^\mathcal{T} \|\widetilde{\vartheta} (t, \omega) \|^2 \,dt
=\infty
$ for $ \mathbb{Q}$-a.e. $ \omega\in E \subseteq\{ \mathcal{T} <
\infty\} $. Then (\ref{C.6w}) implies $ \int_0^\mathcal{T} \|
\vartheta(t, \omega) \|^2 \,dt < \infty$, and $ \widetilde
{\vartheta} (\cdot)= \vartheta(\cdot) - \sigma' (\cdot) \mu
(\cdot)
$ gives $ \int_0^\mathcal{T} \|\widetilde{\vartheta} (t, \omega)
\|^2
\,dt <\infty$, thus $ \mathbb{Q} ( E ) =0 $.
\end{pf}

Equation (\ref{C.6a}) can be thought of as a ``generalized Wald
identity'' [cf. Problem~3.5.7 in Karatzas and Shreve (\citeyear{KS91})]. In Section \ref{sec93} we shall
obtain a characterization of the type (\ref{C.6a}) in a Markovian
context, in terms of properties of an auxiliary diffusion and with the
help of an appropriate partial
differential equation. This will enable us to describe the investment
strategy that realizes the optimal arbitrage.


\subsection{A generalized martingale measure}\label{sec72}

In a similar vein, there exists on the filtered
space $ (\Omega, \mathcal{F}), \mathbb{F} $ a
probability measure $ \widehat{\mathbb{Q}} $ under which
\[
L(t) := 1 / Z(t) = \exp\biggl\{ \int_0^t
\vartheta' (s) \,d \widehat{W}(s) - { 1 \over2} \int_0^t
\| \vartheta(s) \|^2 \,d s \biggr\},\qquad 0 \le t < \infty,
\]
is a martingale, and $ d \mathbb{P}= L (T) \,d \widehat{\mathbb{Q}} $ on each $ \mathcal{F} (T) $, whereas
$ \widehat{W}(\cdot) $ of (\ref{V}) is $ \widehat{\mathbb
{Q}}$-Brownian motion. Under $ \widehat{\mathbb{Q}} $, the processes
$ X_i (\cdot) $, $i=1,\ldots, n $ are nonnegative local (and
super-)martingales, $ d X_i
(t) = X_i (t) \sum_{k=1}^K \sigma_{ik} (t) \,d \widehat{W}_k
(t) $. This justifies the appellation ``generalized
martingale measure'' for $
\widehat{\mathbb{Q}} $.

Defining $ \mathcal{S} := \inf\{ t \ge0 |
L(t)=0 \} $, we have $ \mathbb{P} (\mathcal{S}
<\infty)=0 $ and $ Z(\cdot) $ \textit{is a
strict $ \mathbb{P}$-local martingale if and only if
$ \widehat{\mathbb{Q}}
( \mathcal{S} < \infty) >0 $} [a potential, if and only
if $ \widehat{\mathbb{Q}} ( \mathcal{S} < \infty)
=1 $]; and the expression of (\ref{C.3}), (\ref{C.5}) is
%
%
\begin{equation}
\label{C.6b} \mathfrak{u} (T) =
\mathbb{E}^{\widehat{\mathbb{Q}}} \bigl[
\bigl( X(T) / X(0) \bigr) 1_{\{ \mathcal{S} > T\} } \bigr] .
\end{equation}
This last expression takes the form $
\mathfrak{u}(T)=1-\mathbb{E}^{\widehat{\mathbb{Q}}} [
( X(\mathcal{S}) / X(0) ) 1_{\{ \mathcal{S} \le T\} }
] $ when $ X(\cdot\wedge T) $ is a $
\widehat{\mathbb{Q}}$-martingale; from (\ref{V}), this will be
the case under the Novikov condition $
\mathbb{E}^{\widehat{\mathbb{Q}}} [ \exp\{ { 1
\over2 } \int_0^T \mu' (t) \alpha(t)\mu(t) \,dt \}
] < \infty$. Moreover, $ \mathfrak{u} (T)=1 $ (no
arbitrage relative to the market is possible on $ [0,T] $), if
and only if: $ X(\cdot\wedge T) $ \textit{is a $
\widehat{\mathbb{Q}}$-martingale, and $ X(\mathcal{S}) 1_{\{
\mathcal{S} \le T\}}=0 $ holds $ \widehat{\mathbb{Q}}$-a.e.}

\section{A diffusion model}\label{sec8}

We shall assume from now on that $ K = n $ and that the
processes $ \beta_i (\cdot) $, $ \sigma_{i k}(\cdot) $, $ 1
\le i, k \le n $ in (\ref{1.1}) are of the form
%
%
\begin{equation}
\label{D.1}
\beta_i ( t) = \mathrm{b}_i (\mathfrak{X}(t)) ,\qquad
\sigma_{i k}( t) = \mathrm{s}_{i k} (\mathfrak{X}(t)) ,\qquad
0 \le t < \infty.
\end{equation}
Here $ \mathfrak{X}(t) = ( X_1 (t),\ldots, X_n (t))' $ is the
vector of capitalizations at time $t$, and $ \mathrm{b}_i \dvtx (0,
\infty)^n \rightarrow\mathbb{R} $, $ \mathrm{s}_{i k} \dvtx (0,
\infty)^n \rightarrow\mathbb{R} $ are continuous functions.
We shall denote by $ \mathrm{b} (\cdot) = ( \mathrm{b}_1
(\cdot),\ldots, \mathrm{b}_n (\cdot) )' $ and $ \mathrm{s}
(\cdot) = ( \mathrm{s}_{ik} (\cdot) )_{1 \le i \le n, 1 \le k
\le n} $ the vector and matrix, respectively, of these local
rate-of-return and local volatility functions.
With this setup, the vector process $ \mathfrak{X}(t), 0 \le t
< \infty$ of capitalizations becomes a diffusion, with values
in $ (0, \infty)^n $ and dynamics
%
%
\begin{equation}
\label{D.2}
d X_i (t) = \mathfrak{b}_i (\mathfrak{X}(t)) \,dt +
\sum_{k = 1}^n \mathfrak{s}_{i k} (\mathfrak{X}(t)) \,d W_k (t) ,\qquad
i=1,\ldots, n ,
\end{equation}
where for $ \mathbf{x} = (x_1,\ldots, x_n)' \in(0,
\infty)^n $ we set $ \mathrm{a}_{i j} (\mathbf{x}) := \sum_{k =
1}^n \mathrm{s}_{i k} (\mathbf{x}) \mathrm{s}_{j k} (\mathbf{x})$,
%
%
\begin{equation}
\label{D.3}
\mathfrak{b}_i (\mathrm{x}) := x_i \mathrm{b}_i (\mathbf{x}),\qquad
\mathfrak{s}_{i k} (\mathbf{x}) := x_i \mathrm{s}_{i k}
(\mathbf{x}),\qquad \mathfrak{a}_{i j} (\mathbf{x}) := x_i x_j
\mathrm{a}_{i j} (\mathbf{x}) .
\end{equation}
This diffusion $ \mathfrak{X}(\cdot) $ has infinitesimal generator
%
%
\begin{equation}
\label{D.4}
\mathcal{L} f := {1 \over2 } \sum_{i=1}^n \sum_{j=1}^n
\mathfrak{a}_{ij} (\mathbf{x}) D^2_{ij}f + \sum_{i=1}^n
\mathfrak{b}_{i} (\mathbf{x}) D_{i}f .
\end{equation}
\renewcommand{\theAssumption}{C}
\begin{Assumption}\label{AssumptionC}
For every $ \mathbf{x} \in
(0, \infty)^n $, the matrix $ \mathrm{s}
(\mathbf{x})= ( \mathrm{s}_{ij } (\mathbf{x}) )_{1 \le i, j \le
n} $
is invertible; the system (\ref{D.2}) has a unique-in-distribution weak
solution, with $
\mathfrak{X}(0) = \mathbf{x} $ and values in
$ (0, \infty)^n $;
and for $ \Theta(\mathbf{x}):= \mathrm{s} ^{-1} (\mathbf{x})
\mathrm
{b} (\mathbf{x}) $, the following analogue of
(\ref{1.2}), (\ref{TH}) holds for each $ T \in(0, \infty) $:
%
%
\begin{equation}
\label{D.7}
\sum_{i=1}^n \int_0^T \bigl( | \mathrm{b}_{i}
(\mathfrak
{X}(t) ) | +
\mathrm{a}_{ii} (\mathfrak{X}(t) )+ \Theta_i^2
(\mathfrak
{X}(t) ) \bigr) \,dt <
\infty\qquad \mbox{a.s}.
\end{equation}
\end{Assumption}

It follows from this assumption
that the Brownian motion $ W(\cdot) $ is adapted to the augmentation
of the filtration $ \mathbb{F}^{\mathfrak{X}} $,
and that $ \vartheta(\cdot) = \Theta(
\mathfrak{X}(\cdot) ) $ is a market-price of risk process as
postulated in
(\ref{TH}).
The following conditions from Bass and Perkins (\citeyear{BP03}), in
particular their Theorem 1.2 and Corollary 1.3, are sufficient for
the existence of a weak solution for (\ref{D.2}) which is unique in
distribution:
the functions $ \mathfrak{s}_{i k} (\cdot) $, $ \mathfrak{b}_{i}
(\cdot) $ of (\ref{D.3}) can be extended by
continuity on all of $ [0, \infty)^n $; $ \mathfrak{b}_i
(\cdot) $ and $ \mathfrak{h}_{ij}(\mathbf{x}) := \sqrt{x_i
x_j } \mathrm{a}_{ij} (\mathbf{x}) >0 $ are H\"older
continuous on compact subsets of $ [0,\infty)^n $; and we have
%
%
\begin{eqnarray}
\label{BP}
\mathfrak{b}_i (\mathbf{x}) &\ge& 0 \qquad\mbox{for } x_i=0
;\nonumber\\[-8pt]\\[-8pt]
\| \mathfrak{b} (\mathbf{x})\| + \| \mathfrak{s} (\mathbf{x})\|
&\le&
C ( 1+ \|\mathbf{x}
\| ) \qquad \forall\,\mathbf{x} \in[0,\infty)^n,\nonumber
\end{eqnarray}
and $ \mathfrak{h}_{ij}(\mathbf{x}) =0 $ for $ i \neq
j $, $ \mathbf{x} \in\mathcal{O}^n $, where $
\mathcal{O}^n $ is the boundary of $ [0,\infty)^n $.
\begin{Remark}\label{Remark2}
The diffusion $ \mathfrak{X}(\cdot)$ of (\ref{D.2}) takes values in
$(0, \infty)^n$, if and only if the diffusion $ \Xi(\cdot)= (\Xi_1
(\cdot),\ldots, \Xi_n (\cdot))', \Xi_i (\cdot):= 1 / X_i
(\cdot)
$, with dynamics
%
%
\begin{equation}
\label{Xi}
d \Xi_i (t) = \mathfrak{q}_i (\Xi(t)) \,dt +
\sum_{k = 1}^n \mathfrak{r}_{i k} (\Xi(t)) \,d W_k (t) ,\qquad
i=1,\ldots, n,
\end{equation}
and $
\mathfrak{r}_{i k} (\xi):= - \xi_i \mathrm{s}_{i k} (1 / \xi_1,\ldots, 1 / \xi_n ), \mathfrak{q}_{i } (\xi):= \xi_i (
\mathrm{a}_{i i} -\mathrm{b}_{i } ) (1 / \xi_1,\ldots, 1
/ \xi
_n )$, takes values in $(0, \infty)^n$. Thus, any conditions guaranteeing
the existence of a nonexplosive solution to the SDEs of (\ref{Xi}) for
all times, such as linear growth for $ \mathfrak{q}_{i } (\cdot) $
and $\mathfrak{r}_{i k} (\cdot)$, also ensure that $ \mathfrak
{X}(\cdot)$ takes values in $(0, \infty)^n$.

Alternatively, one may invoke results of Friedman (\citeyear{F06}), Section 9.4
and Chapter 11, to obtain conditions on $ \mathfrak{b}_{i }(\cdot
),
\mathfrak{s}_{i k} (\cdot)$ under which the diffusion $ \mathfrak
{X}(\cdot)$ of (\ref{D.2}) never attains any of the faces $ \{
x_1=0\}
,\ldots, \{ x_n = 0\}$ of $ \mathcal{O}^n $. In particular, if these
functions can be extended by continuity on all of $ [0, \infty)^n $;
the $ \mathfrak{s}_{i k} (\cdot)$ are continuously differentiable;
the matrix $ \mathfrak{a}(\cdot)$ degenerates on the faces of the
orthant; and the so-called \textit{Fichera drifts}
%
%
\begin{equation}
\label{Fichera}
\mathfrak{f}_i (\mathbf{ x}) := \mathfrak{b}_{i } (\mathbf{ x})
- {
1 \over2 } \sum_{j=1}^n D_j \mathfrak{a}_{i j} (\mathbf{ x})
\end{equation}
are nonnegative on $\{ x_i =0\}$, for each $ i=1,\ldots, n $; then
$ \mathfrak{X}(\cdot)$ takes values in $(0, \infty)^n $ [see
Friedman (\citeyear{F06}), Theorem 9.4.1 and Corollary 9.4.2].
\end{Remark}
\renewcommand{\theAssumption}{D}
\begin{Assumption}\label{AssumptionD}
There exists
$ H \dvtx (0, \infty)^n \rightarrow\mathbb{R} $ of class
$ \mathcal{C}^2 $, such that
%
%
\begin{equation}\label{D.6}
\mathfrak{b} (\mathbf{x}) = \mathfrak{a} (\mathbf{x}) D H
(\mathbf{x}), \qquad \forall\,\mathbf{x} \in(0, \infty)^n .
\end{equation}
\end{Assumption}

In light of Assumption \ref{AssumptionC}, this new requirement amounts essentially
to postulating that the vector field $ \mathfrak{a}^{-1}
(\cdot) \mathfrak{b} (\cdot) $ be conservative; it is imposed
here for technical reasons (cf. discussion in Remark \ref{Remark3}). Under it,
the generator of (\ref{D.4}) becomes $ \mathcal{L} f (\mathbf{x}) =
\sum
_{i=1}^n \sum_{j=1}^n \mathfrak{a}_{ij} (\mathbf{x}) [ {1 \over2
} D^2_{ij}
f(\mathbf{x}) + D_i f(\mathbf{x}) D_j H(\mathbf{x})] $, and we have
%
%
\begin{equation}
\label{D.7.a}\quad
\Theta( \mathbf{x} )=\mathfrak{s}' ( \mathbf{x}) D H
( \mathbf{x}) \quad \mbox{and}\quad \mathfrak{s}
(\mathbf{x}) \Theta( \mathbf{x} ) = \mathfrak{b}
(\mathbf{x}) ,\qquad \mathbf{x} \in(0, \infty)^n .
\end{equation}

\textit{Throughout the remainder, Assumptions \ref{AssumptionB},
\ref{AssumptionC}, \ref{AssumptionD} will be in force, and
$ \mathbb{F} \equiv\mathbb{ F }^{\mathfrak{X}}_+ $;
this is a natural choice, and consistent with Assumption \ref{AssumptionA}.}


\section{A parabolic PDE for the function $ U(\tau, \mathbf{x})$}\label{sec9}

The uniqueness in distribution posited in Assumption \ref{AssumptionC} implies
that $ \mathfrak{X}(\cdot) $ is strongly Markovian; we shall
denote by $ \mathbb{P}^{\mathbf{x}} $ the distribution of this
process started at $ \mathfrak{X}(0)=\mathbf{x} \in(0,
\infty)^n $. Our objective now is to study
%
%
\begin{equation}
\label{D.10.a}
U(T, \mathbf{x}) :=
\mathbb{E}^{\mathbb{P}^{\mathbf{x}}} [ Z(T) X(T) ]
/ ( x_1 + \cdots+ x_n ) ,
\end{equation}
the quantity of (\ref{C.3}), (\ref{C.5}) in this diffusion
context.
We start by observing that with $
H(\cdot) $ as in Assumption \ref{AssumptionD} and the notation of (\ref{D.4})
and (\ref{D.7.a}), It\^o's rule gives
$
H (\mathfrak{X}(T) )- H (\mathfrak{X}(0) ) -
\int_0^T \mathcal{L} H (\mathfrak{X}(t) ) \,dt
= \int_0^T \Theta'
(\mathfrak{X}(t) ) \,dW(t) $,
and the exponential local martingale $ Z(\cdot) $ of
(\ref{Z}) becomes
%
%
\begin{equation}
\label{D.9}
Z(\cdot) = \exp\biggl\{ H (\mathfrak{X}(0) )- H
(\mathfrak{X}(\cdot) ) - \int_0^\cdot k
(\mathfrak{X}(t) ) \,dt \biggr\}.
\end{equation}
In particular, $ Z(\cdot) $ is $ \mathbb{F}^\mathfrak{X}$-adapted. We
are setting here
%
%
\begin{eqnarray}\quad
\label{D.10}
k (\mathbf{x}) &:=& - \sum_{i=1}^n \sum_{j=1}^n {
\mathfrak{a}_{ij} (\mathbf{x}) \over2 } [ D^2_{ij}
H(\mathbf{x}) + D_i H(\mathbf{x}) D_j H(\mathbf{x}) ] ,
\\
%
%
\label{D.11}
g(\mathbf{x}) &:=& e^{ -
H(\mathbf{x})} \sum_{i=1}^n x_i ,\qquad G(T, \mathbf{x})
:= \mathbb{E}^{\mathbb{P}^{\mathbf{x}}} \bigl[
g (\mathfrak{X}(T) ) e^{ - \int_0^T k
(\mathfrak{X}(t) ) \,dt } \bigr] .
\end{eqnarray}
With this notation, the function of (\ref{D.10.a}) becomes
$
U(T, \mathbf{x}) = G(T, \mathbf{x}) /
g(\mathbf{x}) $. A~bit more generally, these
considerations---coupled with the Markov property of
$ \mathfrak{X}(\cdot) $---lead for any
$ 0 \le t \le
T $ to the a.s. identity
%
%
\begin{equation}
\label{D.12}
{ \mathbb{E}^{\mathbb{P}^{\mathbf{x}}} [ X(T) Z(T) |
\mathcal{F} (t) ] \over X(t) Z(t) } = { G(T-t,
\mathbf{y}) \over g(\mathbf{y})} \bigg|_{ \mathbf{\mathbf{y}} =
\mathfrak{X} (t)} = U \bigl(T-t,\mathfrak{X} (t) \bigr) .
\end{equation}

The following Assumption \ref{AssumptionE} will also be imposed from now onward.
\textit{It
amounts to assuming that the function $ U ( \cdot, \cdot)
$ of \textup{(\ref{D.10.a})} is of class $ \mathcal{C}^{1,2} $.} Note that
(\ref{D.12.a}) is satisfied, at least in the support of $
\mathfrak{ X} (\cdot)$, thanks to the assumption $ U( \cdot,
\cdot)\in\mathcal{C}^{1,2} ( (0, \infty) \times(0,\infty
)^n
)$ and to the $ \mathbb{P}^\mathbf{ x}$-martingale property of the
process $ G (T-t, \mathfrak{ X} (t) ) e^{- \int_0^t k
(\mathfrak{ X} (u)) \,du } , 0 \le t \le T $.
\renewcommand{\theAssumption}{E}
\begin{Assumption}\label{AssumptionE}
The function
$ G( \cdot, \cdot) $ in (\ref{D.11}) takes values
in $ (0, \infty) $, is continuous on $ [0, \infty) \times(0,
\infty
)^n $, of class $ \mathcal{C}^{1,2} $ on $ (0, \infty) \times(0,
\infty)^n $, and solves
%
%
\begin{eqnarray}
\label{D.12.a}\qquad
{ \partial G \over\partial\tau}
(\tau, \mathbf{x}) &=&
\mathcal{L} G (\tau, \mathbf{x})- k ( \mathbf{x}) G(\tau,
\mathbf{x}),\qquad \tau\in(0, \infty), \mathbf{x} \in(0,
\infty)^n,
\\
%
%
\label{D.12.b}
G (0, \mathbf{x}) &=& g ( \mathbf{x}) ,\qquad \mathbf{x} \in
(0, \infty)^n .
\end{eqnarray}
\end{Assumption}

This Cauchy problem is exactly the one arising in classical
Feynman--Kac theory [see, for instance, Friedman (\citeyear{F06}), Sections 5.6,
6.5,
Karatzas and Shreve (\citeyear{KS91}), Section 5.7 and Janson and Tysk (\citeyear{JT06})].
From Theorem 1 and the remark following it in Heath and Schweizer (\citeyear{HS00}), Assumption \ref{AssumptionE}
holds if: the functions $ \mathfrak{b}_{i } (\cdot)$, $ \mathfrak
{s}_{i k} (\cdot) $ of (\ref{D.3}) are continuously differentiable on
$ (0, \infty) $ and satisfy the growth condition in (\ref{BP}); the
functions $ \mathfrak{a}_{ij} (\cdot)$ of (\ref{D.3}) satisfy the
nondegeneracy condition (\ref{NonDeg}) below; the function $
g(\cdot
) $ in (\ref{D.11}) is H\"older continuous, uniformly on compact
subsets of $ (0, \infty)^n $; the continuous function $ k(\cdot
)
$ of (\ref{D.10}) is bounded from below; and the function $ G (
\cdot, \cdot) $ in (\ref{D.11}) is real-valued and continuous on
$ (0, \infty) \times(0, \infty)^n$. This latter requirement is
satisfied, for instance, if the functions $
\mathfrak{r}_{i k} (\cdot), \mathfrak{q}_{i } (\cdot) $ in
(\ref
{Xi}) obey linear growth conditions, and the function $ \mathfrak{g}
(\xi) := g (1 / \xi_1,\ldots, 1 / \xi_n) $ has polynomial growth [see
Karatzas and Shreve (\citeyear{KS91}), Problem 5.3.15, as well as Heath and
Schweizer (\citeyear{HS00}), Lemma 2 (and the paragraph preceding it)].

Sustained computation shows then that the Cauchy problem of
(\ref{D.12.a}), (\ref{D.12.b}) for $ G( \cdot,
\cdot) $, leads to a corresponding Cauchy problem for
$ U( \cdot, \cdot) $, namely
%
%
\begin{equation}
\label{E.1}\quad
{ \partial U \over\partial\tau} (\tau, \mathbf{x})
= {1 \over2 } \sum_{i=1}^n \sum_{j=1}^n \mathfrak{a}_{ij}
(\mathbf{x}) D^2_{ij} U (\tau, \mathbf{x}) +
\sum_{i=1}^n \sum_{j=1}^n { \mathfrak{a}_{i j}
(\mathbf{x}) D_{i}U
(\tau, \mathbf{x}) \over x_1 + \cdots+ x_n }
\end{equation}
for $ (\tau, \mathbf{x}) \in(0, \infty)
\times(0, \infty)^n $; and
$ U(0 ,\mathbf{x}) = 1 $ for $ \mathbf{x} \in(0, \infty)^n $.


\subsection{\texorpdfstring{An informal derivation of (\protect\ref{E.1})}{An informal
derivation of (9.8)}}\label{sec91} 

Rather than including the computations which lead from
(\ref{D.12.a}) to equation (\ref{E.1}), we present here a
rather simple, informal argument that we shall find useful also in
the next subsection, in a more formal setting. We start by casting
(\ref{C.7.aa}) as
\[
{ d ( X(t) Z(t)) \over X(t) Z(t)}=\sum_{k = 1}^n \Biggl(
\sum_{i=1}^n \mu_i (t) \sigma_{i k} (t) - \vartheta_k (t) \Biggr)
\,d W_k (t)=-\sum_{k = 1}^n \widetilde{\Theta}_k (
\mathfrak{X} (t) ) \,d W_k (t),
\]
where, by analogy with (\ref{C.8}), we have set
%
%
\begin{equation}
\label{D.13}
\widetilde{\Theta}_k (\mathbf{x}) := \Theta_k (\mathbf{x}) -
\sum_{i=1}^n \biggl( { x_i \mathrm{s}_{i k} (\mathbf{x})
\over x_1 + \cdots+ x_n } \biggr),\qquad k = 1,\ldots, n .
\end{equation}
On the other hand, assuming that $ U ( \cdot, \cdot) $ of
(\ref{D.10.a}) is of class $ \mathcal{C}^{1,2} $, we obtain from
It\^
o's rule and with $ R_k (\tau, \mathbf{x}) := \sum_{i=1}^n x_i
\mathrm{s}_{i
k}(\mathbf{x}) D_i U (\tau, \mathbf{x}) $, $ k = 1,\ldots, n $,
\[
d U \bigl(T-t, \mathfrak{X} (t) \bigr) = \biggl( \mathcal{L} U - {
\partial U \over\partial\tau} \biggr) \bigl(T-t,
\mathfrak{X} (t) \bigr) \,dt + \sum_{k = 1}^n R_k \bigl(T-t,
\mathfrak{X} (t) \bigr) \,d W_k (t) .
\]
The product rule of the stochastic calculus applied to the process
%
%
\begin{equation}
\label{D.15}
N(t) := X(t)Z(t) U \bigl(T-t, \mathfrak{X} (t)
\bigr) = \mathbb{E}^{\mathbb{P}^{\mathbf{x}}} [ X(T)
Z(T) | \mathcal{F} (t) ]
\end{equation}
of (\ref{D.12}), leads then to
\begin{eqnarray*}
{ d N(t) \over X(t) Z(t) } &=& d U \bigl(T-t,
\mathfrak{X} (t) \bigr)+ U \bigl(T-t, \mathfrak{X} (t) \bigr)\,{ d (
X(t) Z(t)) \over X(t) Z(t)} \\
&&{} - \sum_{k = 1}^n R_k \bigl(T-t,
\mathfrak{X} (t) \bigr) \widetilde{\Theta}_k ( \mathfrak{X}
(t) ) \,dt
\\
&=& C \bigl(T-t, \mathfrak{X} (t) \bigr) \,dt\\
&&{} + \sum_{k
= 1}^n \bigl[ R_k \bigl(T-t, \mathfrak{X} (t) \bigr)-U \bigl(T-t,
\mathfrak{X} (t) \bigr) \widetilde{\Theta}_k ( \mathfrak{X}
(t) ) \bigr] \,d W_k (t) .
\end{eqnarray*}
We have set
\begin{eqnarray*}
C(\tau, \mathbf{x}) :\!&=& \biggl( \mathcal{L} U - { \partial U
\over\partial\tau} \biggr) (\tau, \mathbf{x}) - \sum_{k =
1}^n R_k (\tau, \mathbf{x}) \widetilde{\Theta}_k ( \mathbf
{x})
\\
&=& {1 \over2 } \sum_{i=1}^n \sum_{j=1}^n \mathfrak{a}_{ij}
(\mathbf{x}) D^2_{ij} U (\tau, \mathbf{x}) +
\sum_{i=1}^n \sum_{j=1}^n { \mathfrak{a}_{i j} (\mathbf{x}) D_{i}U
(\tau, \mathbf{x}) \over x_1 + \cdots+ x_n } - { \partial
U \over\partial\tau}
(\tau, \mathbf{x}),
\end{eqnarray*}
where the last equality is checked easily from
(\ref{D.4}) and (\ref{D.7.a}). But the process $ N(\cdot) $ of
(\ref{D.15}) is a martingale, so the term $
C(\tau, \mathbf{x}) $ should vanish, and
%
%
\begin{equation}
\label{E.0}
{ d N(t) \over N(t) } = \sum_{k = 1}^n \biggl[
{ R_k (T-t, \mathfrak{X} (t) ) \over U (T-t,
\mathfrak
{X} (t) )}
- \widetilde{\Theta}_k ( \mathfrak{X} (t) ) \biggr]
\,dW_k (t) .
\end{equation}
In other words, the function $ U\dvtx [0, \infty) \times(0,
\infty)^n \rightarrow(0,1] $ of (\ref{D.10}) must satisfy the
parabolic partial differential equation (\ref{E.1}), as postulated
earlier.
\begin{Remark}\label{Remark3}
This informal derivation
suggests that it may be possible to dispense with
Assumptions \ref{AssumptionD}, \ref{AssumptionE} altogether, if it can be shown from first
principles that the function
$U$ of (\ref{D.10.a}) is of class $ \mathcal{C}^{1,2} $ on $
(0, \infty) \times(0, \infty)^n $.
Indeed, under suitable conditions, one can rely on techniques from the
Malliavin calculus and the
H\"ormander hypo\"ellipticity theorem [Nualart (\citeyear{N95}),
pages 99--124] to show that the $
(n+2)$-dimensional vector $ ( \mathfrak{X}(T), \Upsilon
(T), \Xi(T) ) $ with $ \Upsilon(T) := \int_0^T \Theta
( \mathfrak{X}(t) )' \,d W(t) $ and $ \Xi(T) :=
\int_0^T \| \Theta( \mathfrak{X}(t) )\|^2 \,dt $ has an
infinitely differentiable probability density function, for any
given $ T \in(0, \infty) $. This provides the requisite
smoothness for the function
\[
U(T, \mathbf{x}) = { 1 \over x_1 + \cdots+ x_n }
\mathbb{E}^{\mathbb{P}^{\mathbf{x}}} \bigl[ \bigl( X_1(T) +
\cdots+ X_n (T) \bigr) e^{ \Upsilon(T) - (\Xi(T)/2
)} \bigr].
\]

The conditions needed for this approach to work are strong;
they include the infinite differentiability of the functions $
\mathfrak{s}_{ik} (\cdot) $, $ \Theta_i (\cdot) $, $1 \le i,k
\le n $, as well as additional algebraic conditions which, in the
present context, are somewhat opaque and not very easy to state or
verify. For these reasons we have opted for sticking with
Assumptions \ref{AssumptionD}, \ref{AssumptionE}; these are satisfied in the Examples of Section \ref{sec12},
are easy to test and allow us to represent
F\"ollmer's exit measure via
(\ref{E.6.u}), (\ref{E.6.v}) without involving stochastic integrals.
\end{Remark}


\subsection{Results and ramifications}\label{sec92}

Equation (\ref{E.1}) is determined \textit{entirely} from the
volatility structure of model (\ref{1.1}). Furthermore, the
Cauchy problem of (\ref{E.1}), $ U (0, \cdot) =1 $, admits the
trivial solution $ U(\tau,\mathbf{x})\equiv1 $; thus, the
existence of arbitrage relative to the market portfolio over a
finite time--horizon $ [0,T] $ is tantamount to \textit{failure
of uniqueness} for the Cauchy problem of (\ref{E.1}), $ U(0,
\cdot) =1 $ over the strip $ [0,T]\times(0, \infty)^n $.
\begin{Remark}\label{Remark4}
Assume there exists some $ h >0 $ such that the continuous
functions $ \mathrm{a}_{ij} (\cdot) ,1 \le i, j \le n $
satisfy either of the conditions
%
%
\begin{eqnarray}\qquad\quad 
\label{E.2.a}
( x_1 + \cdots+ x_n) \sum_{i=1}^n x_i \mathrm{a}_{ii}
(\mathbf{x})- \sum_{i=1}^n \sum_{j=1}^n x_i x_j \mathrm{a}_{ij}
(\mathbf{x}) &\ge& h ( x_1 + \cdots+ x_n)^2 ,
\\
%
%
\label{E.2.abc}
( x_1 \cdots x_n )^{1/n} \Biggl[ \sum_{i=1}^n
\mathrm{a}_{ii} (\mathbf{x})- { 1 \over n } \sum_{i=1}^n
\sum_{j=1}^n \mathrm{a}_{ij} (\mathbf{x}) \Biggr] &\ge& h ( x_1 +
\cdots+ x_n)
\end{eqnarray}
for all $ \mathbf{x} \in(0, \infty)^n $ [we have just re-written
(\ref{IntVol}) and (\ref{EW})
in the present context]. Then from the results
reviewed in Section \ref{sec4} we deduce that, for all $ T
> (2 \log n)/h $ under (\ref{E.2.a}), and for all $ T
> (2 n^{1-(1/n)} )/h $ under (\ref{E.2.abc}), we
have $
U (T, \mathbf{x}) < 1 , \forall\,\mathbf{x}
\in(0, \infty)^n $. In particular, under either (\ref{E.2.a})
or (\ref{E.2.abc}), \textit{uniqueness fails for the Cauchy problem
of} (\ref{E.1}), $ U(0, \cdot) \equiv1 $.

Whenever uniqueness fails for this problem, it is important to know
how to pick the ``right'' solution from among all possible solutions,
the one which gives
the quantity of (\ref{D.10.a}). The next result addresses this issue;
it implies that $ G (\cdot, \cdot) $ in (\ref{D.11}) is the
\textit{smallest nonnegative}, continuous function, of class $ \mathcal
{C}^{1,2} ( (0, \infty) \times(0, \infty)^n ) $, which
satisfies $ ( \partial G / \partial\tau) \ge\mathcal{L} G - k G
$ and (\ref{D.12.b}) [cf. Karatzas and Shreve (\citeyear{KS91}), Exercise 4.4.7 for
a similar situation].
\end{Remark}
\begin{Theorem}\label{Theorem1}
The function $ U\dvtx [0, \infty) \times(0, \infty)^n \rightarrow(0,1]
$ of (\ref{D.10.a}) is the \textup{smallest nonnegative} continuous
function, of class $ \mathcal{C}^{1,2} $ on $ (0, \infty) \times(0,
\infty)^n $, that satisfies $ U(0, \cdot) \equiv 1 $ and (\ref{E.3}).
\end{Theorem}
\begin{pf}
Consider any continuous function $ \widetilde{U}\dvtx [0,
\infty) \times(0, \infty)^n \rightarrow[0,\infty) $ which is of
class $ \mathcal{C}^{1,2} $ on $ (0,\infty) \times(0, \infty
)^n
$, and satisfies (\ref{E.3}) and $ \widetilde{U}(0, \cdot) \equiv
1
$ on $ (0, \infty)^n$; we shall denote by $ \mathfrak{ U} $ the
collection of all such functions. We introduce $ \widetilde{N}(t) :=
X(t)Z(t)\widetilde{U} (T-t, \mathfrak{X} (t) ) $, $ 0
\le
t \le T $ as in (\ref{D.15}).

Repeating verbatim the arguments in Section \ref{sec91}, we use
(\ref{E.3}) to conclude that the nonnegative process $
\widetilde{N}(\cdot) $ is a
local supermartingale. Thus $
\widetilde{N}(\cdot) $ is bona-fide supermartingale, $
(x_1 + \cdots+ x_n) \widetilde{U} (T,
\mathbf{x})= \widetilde{N}(0) \ge
\mathbb{E}^{\mathbb{P}^{\mathbf{x}}} ( \widetilde{N}(T) )=
\mathbb{E}^{\mathbb{P}^{\mathbf{x}}} ( X(T) Z(T) ) $ holds
for every
$ (T,\mathbf{x}) \in(0, \infty) \times(0, \infty)^n $, and $
\widetilde{U} (T, \mathbf{x}) \ge U (T, \mathbf{x}) $ follows
from (\ref{D.10.a}).
\end{pf}
\begin{Proposition}\label{Proposition2}
Assume that the continuous
functions $ ( \mathfrak{a}_{ij} (\cdot) )_{ 1 \le i, j
\le n} $ of (\ref{D.3}) satisfy the following nondegeneracy
condition: \textit{for every
compact subset
$ \mathcal{K} $ of $ (0,\infty)^n $, there exists a number $
\varepsilon= \varepsilon_\mathcal{K}
>0 $ such that}
%
%
\begin{equation}
\label{NonDeg}
\sum_{i=1}^n \sum_{j=1}^n \mathfrak{a}_{ij} (\mathbf{z}) \xi_i
\xi_j \ge
\varepsilon\| \xi\|^2, \qquad \forall\,\mathbf{z} \in
\mathcal{K} , \xi\in\mathbb{R}^n .
\end{equation}

Then, if
%
%
\begin{equation}
\label{E.3.u1}
U (T, \mathbf{x}) < 1 \qquad\mbox{\textit{for some} }
\mathbf{x} \in
(0,\infty)^n
\end{equation}
holds for some $ T \in(0,\infty) $,
we have
%
%
\begin{equation}
\label{E.3.v}
U (T, \mathbf{x}) < 1, \qquad \forall(T, \mathbf{x}) \in
(0,\infty) \times(0,\infty)^n .
\end{equation}
\end{Proposition}
\begin{pf}
Let us work first under the stronger assumption
%
%
\begin{equation}
\label{E.3.u}
U (T, \mathbf{x}) < 1, \qquad \forall\,
\mathbf{x} \in(0,\infty)^n ,
\end{equation}
for some $ T \in(0,\infty) $. For every $ \tau>0 $, we
consider the set $
\mathcal{S} (\tau) := \{ \mathbf{x} \in(0,\infty)^n | U(\tau,
\mathbf{x})=1\} $ and define $ \tau_* := \sup\{ \tau
\in(0,\infty) | \mathcal{S} (\tau) \neq\varnothing\} $ (with
$ \tau_*=0 $ if the set is empty). Assumption
(\ref{E.3.u}) amounts to $ \tau_* <\infty$,
and the claim (\ref{E.3.v}) to $ \tau_*=0 $; we shall prove this
claim by contradiction.

Suppose $ \tau_*>0 $; then $ U (\tau_* - \delta,
\mathbf{x}_*)=1 $ for any given $ \delta\in(0, \tau_*/2) $,
and some $ \mathbf{x}_*\in(0,\infty)^n $. For \textit{any} given
$ \mathbf{x}\in(0,\infty)^n $, consider an open, connected set
$ D $ which contains both $ \mathbf{x} $ and $
\mathbf{x}_* $, and whose closure $ \overline{D} $ is a
compact subset of $ (0,\infty)^n $; in particular, we have $
\inf\{ \| \mathbf{y} - \mathbf{z}\| | \mathbf{z} \in
\overline{D} , \mathbf{y} \in\mathcal{O}^n \}>0 $. The
function $ U( \cdot, \cdot) $ attains its maximum value
over the cylindrical domain $ \mathfrak{E} = \{ (\tau
, \mathbf{\xi} ) | 0 < \tau< \tau_* +1 ,
\mathbf{\xi} \in D \} $ at the point $ (\tau_* -
\delta, \mathbf{x}_*) $, which lies in the interior of this
domain. By assumption then, the operator
$
\widehat{\mathcal{L}} f =(1 /2) \sum_{i=1}^n \sum_{j=1}^n
\mathfrak{a}_{ij} (\mathbf{x}) D^2_{ij}f + \sum_{i=1}^n
\widehat{\mathfrak{b}}_{i} (\mathbf{x}) D_{i}f $
of (\ref{E.3.a}) with
%
%
\begin{equation}
\label{E.3.b}\qquad
\widehat{\mathfrak{b}}_{i} (\mathbf{x}) := x_i
\widehat{\mathrm{b}}_i (\mathbf{x}) ,\qquad
\widehat{\mathrm{b}}_i(\mathbf{x}) := \sum_{j=1}^n {x_j
\mathrm{a}_{i j} (\mathbf{x}) \over x_1 + \cdots+
x_n } ,\qquad i=1,\ldots, n ,
\end{equation}
is uniformly parabolic with bounded, continuous
coefficients on $ \mathfrak{E} $, so from the maximum
principle for parabolic operators [Friedman (\citeyear{F06}), Chapter 6],
%
%
\begin{equation}
\label{E.3.g}
U (\tau, \mathbf{x}) = 1 \qquad \forall( \tau, \mathbf{x})
\in[0,\tau_* - \delta) \times(0,\infty)^n .
\end{equation}
Now let us recall the $ \mathbb{P}^{\mathbf{x}}$-a.s. equality
$
\mathbb{E}^{\mathbb{P}^{\mathbf{x}}} [ X(T) Z(T) |
\mathcal{F} (t) ]
= U (T-t,\mathfrak{X} (t) )\cdot X(t) Z(t)
$ from (\ref{D.12}); we apply it with $ 0\le t \le\tau_*-
\delta$, $ 0\le T-t \le\tau_*- \delta$, then take
expectations with respect to the probability measure
$ \mathbb{P}^{\mathbf{x}} $, and
use (\ref{E.3.g}) along with (\ref{D.10.a}), to obtain
for every $ T \in[ 0, 2 (\tau_*- \delta) ] $,
\[
U(T,\mathbf{x})={ \mathbb{E}^{\mathbb{P}^{\mathbf{x}}} [
X(T) Z(T) ] \over x_1 + \cdots+ x_n} ={
\mathbb{E}^{\mathbb{P}^{\mathbf{x}}} [ X(t) Z(t) ]
\over x_1 + \cdots+ x_n}= U(t,\mathbf{x})=1, \qquad \forall\,
\mathbf{x} \in(0,\infty)^n.
\]
But since $ 2 (\tau_*- \delta)
>\tau_* $, this contradicts the definition
of $ \tau_* $.

Now we revert to (\ref{E.3.u1}); as J. Ruf (private communication)
observes, yet another application of the maximum principle, as above,
leads to (\ref{E.3.u}).
\end{pf}
\begin{Corollary*} Under the nondegeneracy
condition (\ref{NonDeg}), and with either (\ref{E.2.a}) or
(\ref{E.2.abc}), inequality (\ref{E.3.v}) holds.
That is, arbitrage with respect to the market exists then over
\textup{any} time--horizon $[0,T]$ with $ T \in(0,\infty)$.
\end{Corollary*}


\subsection{An auxiliary diffusion}\label{sec93}

Let us consider now the diffusion process $
\mathfrak{Y}(\cdot) $ with infinitesimal generator $ \widehat
{\mathcal{L}} $ as in
(\ref{E.3.a}), (\ref{E.3.b}) and dynamics
%
%
\begin{equation}
\label{E.4}
d Y_i (t) = \widehat{\mathfrak{b}}_i (\mathfrak{Y}
(t) ) \,dt + \sum_{k = 1}^n \mathfrak{s}_{i k}
(\mathfrak{Y}(t) ) \,d W_k (t) ,\qquad i=1,\ldots, n.
\end{equation}
\renewcommand{\theAssumption}{F}
\begin{Assumption}\label{AssumptionF}
The system of SDEs
(\ref{E.4}) admits a unique-in-distribution weak solution with
values in $ [0, \infty)^n \setminus\{ \mathbf{0} \} $.
\end{Assumption}

This will be the case, for instance, if the drift functions $
\widehat{\mathfrak{b}}_i (\cdot) , 1 \le i \le n $ of (\ref
{E.3.b}) can be extended by continuity on all of $ [0, \infty)^n $
and satisfy the Bass and Perkins (\citeyear{BP03}) conditions preceding, following
and including (\ref{BP}). The resulting process $ \mathfrak{Y}(\cdot
)
$ is then Markovian, and we shall denote by $
\mathbb{Q}^{\mathbf{y}} $ its distribution
with $ \mathfrak{Y}(0)=\mathbf{y} \in[0, \infty)^n $. Unlike
the original process $ \mathfrak{X}(\cdot) $, which takes
values in $ (0, \infty)^n $, this new process
$ \mathfrak{Y}(\cdot) $ is only guaranteed to take values in the
nonnegative orthant $ [0, \infty)^n \setminus\{ \mathbf{0}
\} $. In particular, with $ \mathbf{x} \in(0, \infty)^n $ the
first hitting time
%
%
\begin{equation}
\label{E.5}
\mathfrak{T} := \inf\{ t \ge0 | \mathfrak{Y}(t) \in
\mathcal{O}^n \}
\end{equation}
of the boundary $ \mathcal{O}^n $ of $ [0, \infty)^n $ may
be finite with positive $ \mathbb{Q}^{\mathbf{x}}$-probability.

Our next result shows that this possibility amounts to the
existence of arbitrage relative to the market, and to the lack of
uniqueness for the Cauchy problem of (\ref{E.1}) and $ U(0,
\cdot) \equiv1 $.
\begin{Theorem}\label{Theorem2}
With the above notation and assumptions,
including (\ref{NonDeg}), the function $ U\dvtx [0, \infty) \times(0,
\infty)^n \rightarrow(0,1] $ of (\ref{D.10.a}) admits the representation
%
%
\begin{equation}
\label{E.6}
U(T,\mathbf{x}) = \mathbb{Q}^{\mathbf{x}} [ \mathfrak{T}
>T ] ,\qquad (T,\mathbf{x}) \in(0, \infty) \times(0, \infty
)^n .
\end{equation}
\end{Theorem}
\begin{pf}
The function on the right-hand side of (\ref{E.6}) is
space--time harmonic for the diffusion $ \mathfrak{Y} (\cdot) $
on $(0, \infty) \times(0, \infty)^n$, so it solves equation
(\ref{E.1}) there [cf. Janson
and Tysk (\citeyear{JT06}),  Theorem 2.7]. Consider any function $ V $ in
the collection $ \mathfrak{ U} $ of Theorem \ref{Theorem1}; then $ V(T-t,
\mathfrak{Y}(t)) 1_{\{\mathfrak{T}
>t \}} , 0 \le t \le T $ is a nonnegative local (thus a true) $
\mathbb{Q}^{\mathbf{x}}$-supermartingale, and we deduce
\begin{eqnarray}
V(T,\mathbf{x}) \ge
\mathbb{E}^{\mathbb{Q}^{\mathbf{x}}} \bigl[
V (0, \mathfrak{Y}(T ) ) 1_{\{\mathfrak{T}
>T \}} \bigr] = \mathbb{Q}^{\mathbf{x}}
( \mathfrak{T}
>T) ,\nonumber\\
\eqntext{(T,\mathbf{x}) \in(0, \infty) \times
(0, \infty)^n.}
\end{eqnarray}
The claim follows now from the proof of Theorem \ref{Theorem1}.
\end{pf}
\begin{Corollary*}
Under the assumptions of Theorem \ref{Theorem2}, for any
given $ \mathbf{x} \in(0, \infty)^n $ the
$ \mathbb{P}^\mathbf{x}$-supermartingale $ Z(\cdot) X(\cdot) $
is under $ \mathbb{P}^\mathbf{x} $ a:

$\bullet$ martingale, if and only if
$ \mathbb{Q}^\mathbf{x} ( \mathfrak{T} < \infty)=0 $;\vspace*{1pt}

$\bullet$ potential [i.e., $ \lim_{T
\rightarrow\infty} \mathbb{E}^{\mathbb{P}^\mathbf{x}} (
Z(T)X(T) )=0 $], iff $ \mathbb{Q}^\mathbf{x} (
\mathfrak{T} < \infty)=1 $;

$\bullet$ strict local (and super-)martingale
on any time--horizon $ [0,T] $ with $ T \in(0, \infty) $,
if and only if
$ \mathbb{Q}^\mathbf{x}
( \mathfrak{T} < \infty)>0 $.
\end{Corollary*}

We represent by analogy with (\ref{C.6a}) the exit measure $
\mathfrak{P}^\mathbf{x} $ of the supermartingale $ Z(\cdot)
X(\cdot) $ with initial configuration $ \mathfrak{X} (0) =
\mathbf{x} $, in the form
%
%
\begin{equation}
\label{E.6.u}
\mathfrak{P}^\mathbf{x} \bigl( (T, \infty] \times\Omega\bigr)
= U(T,\mathbf{x}) = \mathbb{Q}^{\mathbf{x}} [
\mathfrak{T}>T ],
\end{equation}
and from (\ref{D.9})--(\ref{D.12}) we have
for $ A \in\mathcal{F} (t) , 0 \le t \le T $,
%
%
\begin{eqnarray}
\label{E.6.v}
&&\mathfrak{P}^\mathbf{x} \bigl( (T, \infty] \times A \bigr)\nonumber\\[-8pt]\\[-8pt]
&&\qquad=
\mathbb{E}^{\mathbb{P}^{\mathbf{x}}} \biggl[ {
g (\mathfrak{X}(t) )
\over g(\mathbf{x}) } 1_A \bigl(
\mathbb{Q}^{\mathbf{z}} [ \mathfrak{T}>T-t ] \bigr)
\big|_{\mathbf{z} = \mathfrak{X}(t)}\, e^{ - \int_0^t k
(\mathfrak{X}(s) ) \,ds } \biggr].\nonumber
\end{eqnarray}

When $ \mathbf{x} \in(0, \infty)^n $ and the
quantity of (\ref{E.6}) is equal to one, the
$ \mathbb{Q}^{\mathbf{x}}$-distribution of the process $
\mathfrak{Y}( t), 0 \le t \le T $ in (\ref{E.4}) is the same
as the $ \widetilde{\mathbb{P}}^{\mathbf{x}}_T$-distribution of
the original stock-price process $ \mathfrak{X}( t), 0 \le t
\le T $; this follows\vspace*{1pt} by comparing (\ref{E.4}) and (\ref{E.3.b})
with (\ref{C.9}), and denoting by
$ \widetilde{\mathbb{P}}^{\mathbf{x}}_T $ the probability
measure $ \widetilde{\mathbb{P}}_T $ of (\ref{C.9.a}) with $
\mathfrak{X} (0) = \mathbf{x} $. We have in this spirit the following
result, by analogy with Remark \ref{Remark2}.
\begin{Proposition}\label{Proposition3}
Under the assumptions of Theorem \ref{Theorem2}, suppose
that the functions $ \mathfrak{s}_{i k} (\cdot)$ are continuously
differentiable on $(0, \infty)^n$; that the matrix $ \mathfrak
{a}(\cdot)$ degenerates on $ \mathcal{O}^n$; and that the analogues of
(\ref{Fichera}), the Fichera drifts
%
%
\begin{equation}
\label{FicheraToo}\hspace*{28pt}
\widehat{\mathfrak{f}}_i (\mathbf{x}) := \widehat{\mathfrak{b}}_{i }
(\mathbf{x}) - { 1 \over2 } \sum_{j=1}^n D_j \mathfrak{a}_{i j}
(\mathbf{x})
= \sum_{j=1}^n \biggl( {\mathfrak{a}_{i j} (\mathbf{ x}) \over
x_1 + \cdots+
x_n} - { 1 \over2 } D_j \mathfrak{a}_{i j} (\mathbf{ x})
\biggr)
\end{equation}
for the process $ \mathfrak{ Y} (\cdot)$ of (\ref{E.4}), can be
extended by continuity on $ [0, \infty)^n$.
If $\, \widehat{\mathfrak{f}}_i (\cdot) \ge0 $ holds on each face
$\{ x_i =0 \} $, $ i =1,\ldots, n $ of the orthant, then we have
$ U (\cdot, \cdot) \equiv1 $ in (\ref{E.6}), and no arbitrage
with respect to the market portfolio exists on any time--horizon.

If, on the other hand, we have $ \widehat{\mathfrak{f}}_i (\cdot) <
0 $ on each face $\{ x_i =0 \} $ of the orthant, then $ U (\cdot,
\cdot) < 1 $ in (\ref{E.6}) and arbitrage with respect to the market
portfolio exists, on every time--horizon $ [0, T] $ with $ T \in
(0, \infty)$.
\end{Proposition}
\begin{pf}
In light of Theorem \ref{Theorem2}, the first claim follows from
Theorem 9.4.1, Corollary 9.4.2 of Friedman (\citeyear{F06}), and the second is a
consequence of the support theorem for diffusions [Ikeda and Watanabe
(\citeyear{IW89}), Section VI.8].
\end{pf}
\begin{Remark}\label{Remark5}
(i) The ``relative weights'' $ \nu_i (t) := Y_i (t) /
( Y_1
(t)+ \cdots+ Y_n (t)) $, $ i = 1 ,\ldots, n $ have dynamics
similar to (\ref{2.2.a}),
%
%
\begin{equation}
\label{NUs}
d \nu_i (t) = \nu_i (t) \bigl( \mathfrak{e}_i - \nu(t) \bigr)'
\mathrm{s} ( \mathfrak{Y} (t) )
\,d W (t) .
\end{equation}
They are thus $ \mathbb{Q}^\mathbf{x}$-martingales with
values in $[0,1]$ (cf. Section \ref{sec61}); so, when any one of
them hits either boundary point of the unit interval, it gets
absorbed there. In terms of them, the first hitting time of
(\ref{E.5}) can be expressed as in (\ref{C.6aa}), $
\mathfrak{T} = \min_{1 \le i \le n} \mathfrak{T}_i $, where
$ \mathfrak{T}_i := \inf\{ t \ge0 | \nu_i (t) =0 \}$.

(ii) The measure $ \mathbb{Q}^{\mathbf{x}} $ corresponds to a
\textit{change of drift}, from $ \mathfrak{b} (\cdot) $ in
(\ref{D.2}) to $ \widehat{\mathfrak{b}} (\cdot) $ in
(\ref{E.3.b}), (\ref{E.4}); this ensures that, under $ \mathbb
{Q}^{\mathbf{x}} $, the components of the new, ``fictitious'' market
portfolio $ \nu(\cdot) $ are martingales, that $ \nu(\cdot) $
has the num\'eraire property, and thus that $ \nu(\cdot) $
cannot be outperformed.
\end{Remark}


\section{Markovian market weights}\label{sec10}

Let us assume now
the form
\[
\mathrm{b}_{i} (\mathbf{x})= \mathfrak{B}_{i} ( x_1 / x
,\ldots, x_n / x ) ,\qquad
\mathrm{s}_{i k} (\mathbf{x})= \mathfrak{S}_{i k} ( x_1 / x
,\ldots, x_n / x )
\]
for the functions of
(\ref{D.1}), with $ x := \sum_{j=1}^n x_j $ and suitable continuous
functions $ \mathfrak{B}_{i} (\cdot) $, $
\mathfrak{S}_{i k} (\cdot) $ on $ \Delta^n_+ $. For
$ \mathrm{m} = (m_1,\ldots, m_n)' \in\Delta^n_+ $, we set $
\mathcal{A}_{i j} (\mathrm{m}):= \sum_{k=1}^n \mathfrak{S}_{i k}
(\mathrm{m}) \mathfrak{S}_{j k} (\mathrm{m})$. In words, we consider
instantaneous growth rates and volatilities that depend at time
$ t $ only on the current configuration $ \mu( t) = (
\mu_1 ( t),\ldots, \mu_n ( t) )' $ of relative market
weights, so the process $ \mu(\cdot) $ of (\ref{2.2}) is now
a diffusion with values in the positive simplex $ \Delta^n_+ $
and
%
%
\begin{equation}
\label{G.2}
d \mu_i (t)= \mu_i (t) \Biggl[ \Gamma_i ( \mu(t) ) \,dt +
\sum_{k=1}^n \mathcal{T}_{i k} ( \mu(t) ) \,d W_k (t) \Biggr] ,\qquad i
= 1,\ldots, n ,\hspace*{-36pt}
\end{equation}
with $ \mathcal{T}_{i k} (\mathrm{m}) := \mathfrak{S}_{i k}
(\mathrm{m}) - \sum_{j=1}^n m_j \mathfrak{S}_{j k}
(\mathrm{m}) , \mathcal{P}_{i j} (\mathrm{m}) :=
\sum_{k=1}^n \mathcal{T}_{i k} (\mathrm{m}) \mathcal{T}_{j k}
(\mathrm{m}) , $
\[
\Gamma_i (\mathrm{m}) := \mathfrak{B}_{i } (\mathrm{m}) -
\sum_{j=1}^n m_j \mathfrak{B}_{j } (\mathrm{m}) -
\sum_{j=1}^n m_j
\mathcal{A}_{i j} (\mathrm{m}) + \sum_{j=1}^n \sum_{k=1}^n
m_j m_\ell\mathcal{A}_{j \ell} (\mathrm{m}) .
\]

In this setup, the function of (\ref{D.10.a}) can be expressed in the form
$ U (T, \mathbf{x})= Q ( T, x_1 / x ,\ldots, x_n / x
)$,
in terms of a function $ Q \dvtx (0, \infty) \times\Delta^n_+
\rightarrow(0,1] $ that satisfies the initial condition $ Q (0
, \cdot) \equiv1 $ and the equation
\[
{ \partial Q \over\partial\tau} (\tau, \mathrm{m})
= {1 \over2 } \sum_{i=1}^n \sum_{j=1}^n m_i m_j
\mathcal{P}_{ij} (\mathrm{m}) D^2_{ij} Q
(\tau, \mathrm{m}) ,\qquad (\tau, \mathrm{m})
\in(0, \infty) \times\Delta^n_+ ,
\]
which appears
on page 56 of Fernholz (\citeyear{F02}) and can be derived from
(\ref{E.1}). On the other hand, by analogy with Theorem \ref{Theorem2} and (\ref
{NUs}), the
quantity $ Q (T,\mathrm{m}) $ is the probability that the
process $ \nu(\cdot) = ( \nu_1 (\cdot),\ldots, \nu_n
(\cdot) )' $ with $ \nu(0) = \mathrm{m} \in
\Delta^n_+ $ and dynamics (\ref{G.7}) below, does not hit the
boundary of the nonnegative simplex $ \Delta^n := \{
\mathrm{m} \in[0,1]^n |  \sum_{i=1}^n m_i = 1 \} $
before $ t=T $:
%
%
\begin{equation}
\label{G.7}
d \nu_i (t) = \nu_i (t) \sum_{k=1}^n \mathcal{T}_{i k} (
\nu(t) ) \,d W_k (t),\qquad i = 1,\ldots, n .
\end{equation}


\section{The investment strategy}\label{sec11}

Let us substitute now the expressions of (\ref{D.13}) into
(\ref{E.0}), to obtain the dynamics of the martingale $ N(\cdot)
\equiv Z(\cdot) X(\cdot) U (T-\cdot, \mathfrak{X}(\cdot)
) $ in (\ref{D.15}), with $ N(0)= \xi:= X (0) U(T,
\mathfrak{X}(0)) $,
\begin{eqnarray*}
N(t) &=& \xi+ \sum_{k = 1}^n \int_0^t N(s)
\Psi_k \bigl(T-s, \mathfrak{X} (s) \bigr) \,d W_k (s) ,\qquad
0 \le t \le T ,
\\
\Psi_k (\tau, \mathbf{x}) :\!&=& \sum_{i=1}^n \mathrm{s}_{i k}
(\mathbf{x}) \biggl( x_i D_i \log U (\tau, \mathbf{x}) + { x_i
\over x_1 + \cdots+ x_n } \biggr)-\Theta_k (\mathbf{x}) .
\end{eqnarray*}
Thus we can identify the ``replicating strategy'' $ \widehat{\pi}
(\cdot) $ of (\ref{C.5}) as
%
%
\begin{equation}
\label{F.6}\hspace*{32pt}
\widehat{\pi}_i (t)=X_i (t) D_i \log U \bigl( T-t,
\mathfrak{X}(t) \bigr) + \bigl( X_i (t) / X(t) \bigr),\qquad
i = 1,\ldots, n,
\end{equation}
and its value as $ V^{ \xi,\widehat{\pi} } (t) = N(t) / Z(t)
=X(t) U ( T-t,
\mathfrak{X}(t) ) $, $ 0 \le t \le T $.
\begin{Remark}\label{Remark6}
In the special case of a Markovian
model (\ref{G.2}) for the market weights of $ \mu(\cdot) = ( \mu_1
(\cdot),\ldots, \mu_n (\cdot))' $, expression
(\ref{F.6}) takes the form
\[
\widehat{\pi}_i (t) = \mu_i (t) \Biggl( 1 +
D_i \log Q \bigl( T-t,
\mu(t) \bigr) - \sum_{j=1}^n \mu_j (t) D_j \log Q
\bigl( T-t, \mu(t) \bigr) \Biggr)
\]
of a ``functionally-generated portfolio'' in the terminology of
Fernholz (\citeyear{F02}), page~56; whereas the value
is $
V^{ \xi,\widehat{\pi} } (t) = X(t) Q ( T-t,
\mu(t) ) $, $ 0 \le t \le T $.

In this
case we have $ \sum_{i=1}^n \widehat{\pi}_i (\cdot) \equiv1 $:
the strategy that implements the best possible arbitrage relative
to the equity market never borrows or lends.
\end{Remark}


\section{Examples}\label{sec12}

We discuss in this section two illustrative examples. Additional
examples, in which the investment strategy $ \widehat{\pi} (\cdot
) $
of (\ref{F.6}) that realizes the optimal arbitrage can be computed in
closed form in dimension $ n=1 $, can be found in Ruf
(\citeyear{R09}).\vspace*{9pt}

For the first of these examples, take $n=1$, $ \beta(t) = 1 / X^2
(t) $ and $ \sigma(t) = 1 / X (t) $ in (\ref{1.1}) where the
process $
X(\cdot) $ satisfies $ d X(t) = ( 1 / X(t)) \,dt + d W(t) $ and
$ X(0)=1 $. This is a Bessel process in dimension three---the
radial part of
a \mbox{3-D} Brownian motion started at unit distance from the
origin---and takes values in $ (0, \infty) $. We have then $ \vartheta
(t) = 1 / X(t) $, $ Z (t) = 1 / X(t) $ for $ 0 \le t
<\infty$ in (\ref{TH}) and (\ref{Z}), so $ Z(\cdot) X(\cdot) $
is very clearly a martingale. However, $ Z(\cdot) $ is the
prototypical example of a \textit{strict} local martingale---we have
$ \mathbb{E} (Z(T) ) <1 $ for every $ T \in(0, \infty) $
[e.g., Karatzas and Shreve
(\citeyear{KS91}), Exercise 3.36, page 168]. This example is taken from Karatzas
and Kardaras [(\citeyear{KK07}),
page 469], where an arbitrage with respect to the money-market
is constructed in closed form. It illustrates that it is
possible for $ Z(\cdot) $ to be a strict
local martingale and $ Z(\cdot) X(\cdot) $ to be a martingale; in
other words, the second and third inequalities in
(\ref{AO.b}) fail, while the first stands.

Here we have $ \Theta(x) = 1/x $, $ H(x) =
\log x $ and $ k(\cdot) \equiv0 $, $ g(\cdot) \equiv1 $,
$ G( \cdot, \cdot) \equiv1 $ in (\ref{D.10}),
(\ref{D.11}), thus $ U (T, x) \equiv1 $ for all $ T \in[0,
\infty) $, $ x \in(0, \infty) $. Arbitrage relative to $
X(\cdot
) $ does not exist here, despite the existence of arbitrage relative to
the money market and the fact that $ Z(\cdot) $ is a strict
local martingale. Note that $ \widehat{\mathfrak{b}} (x) =1/x $ in
(\ref{E.3.b}), so the diffusion of (\ref{E.4}) is again a Bessel
process in dimension three, $
d Y (t)= ( 1 / Y(t) ) \,dt + d W(t) $, $ Y(0)=y>0 $. This
process never hits the origin, so the probability in (\ref{E.6}) is
equal to one, for all $ T \in[0, \infty) $.

\subsection{The volatility-stabilized model}\label{sec121}

Our second example is the model of ``stabilization by volatility''
introduced in Fernholz and Karatzas (\citeyear{FK05}) and studied further by
Goia (\citeyear{G09}). With $ n \ge2 $, $ \zeta\in[0,1] $
this posits
%
%
\begin{eqnarray}
\label{D.8*a}
\beta_i (t) &=& ( 1 + \zeta) / (2 \mu_i (t) ),\nonumber\\[-8pt]\\[-8pt]
\sigma_{i k} (t) &=&
\delta_{i k} ( \mu_i (t) )^{-1/2} ;\qquad 1 \le i, k
\le n ,\nonumber
\end{eqnarray}
that is, rates of return and volatilities which
are large for the small stocks and small for the large stocks.
The conditions of Bass and Perkins (\citeyear{BP03})
hold for the resulting system of SDEs in the notation of
(\ref{2.1}) with $\kappa:= (1 +\zeta)/2 $,
%
%
\begin{equation}
\label{D.8*}
d X_i (t) = \kappa X (t) \,dt
+ \sqrt{ X_i (t) X(t) } \,d W_i (t),\qquad
i=1,\ldots, n .
\end{equation}
The unique-in-distribution solution of (\ref{D.8*}) is expressed
in terms of independent Bessel processes $ \mathfrak{R}_1
(\cdot),\ldots, \mathfrak{R}_n (\cdot) $ in dimension $ 4
\kappa
$ with $ X_i (t) = \mathfrak{R}_i^2 (A(t) )
>0 $ and $ A(t) := ( 1 / 4) \int_0^t X(s) \,ds $. In
particular, $ \mathfrak{X}(\cdot) $ takes values in $ (0,
\infty)^n $; for more details on these Lamperti-like
descriptions and their
implications, see Fernholz and Karatzas
(\citeyear{FK05}) and Goia (\citeyear{G09}). Condition (\ref{D.7}) is satisfied in
this example, so Assumption \ref{AssumptionC} also holds.

For the model of (\ref{D.8*a}), we have
$ \Theta_i (\mathbf{x}) / \kappa=
\mathrm{s}_{ii}(\mathbf{x}) = ( (x_1 + \cdots+x_n ) / x_i )^{1/2}$,
\[
\mathfrak{b}_i (\mathbf{x})= \kappa(
x_1 + \cdots+ x_n ) ,\qquad \mathfrak{h}_{i j} (\mathbf{x}) =
\delta_{i j} ( x_1 + \cdots+ x_n ) ,\qquad
\mathfrak{a}_{i j} (\mathbf{x}) = x_i \mathfrak{h}_{i j}
(\mathbf{x})
\]
for $ 1 \le i,j \le n $. The assumptions of Theorem \ref{Theorem2} and of
Propositions \ref{Proposition1} and \ref{Proposition3} are all satisfied here, as are (\ref{C.6w}) and
(\ref{D.6}) with $
H(\mathbf{x})= \kappa\sum_{i=1}^n \log x_i
$ and $ k (\mathbf{x}) = ( 1-\zeta^2 )
(x_1 + \cdots+ x_n ) \sum_{j=1}^n
( 1 / (8 x_j) ) $. This function $ k (\cdot) $ is
nonnegative, since we have assumed $ 0 \le\zeta\le1 $, whereas
$ g (\mathbf{x}) = (x_1
+ \cdots+ x_n) ( x_1 \cdots x_n
)^{-\kappa} $. In particular, with $ \zeta= 1 $ we get
%
%
\begin{equation}
\label{E.U}
U(T, \mathbf{x})
= { x_1 \cdots x_n \over x_1 + \cdots+ x_n }
\mathbb{E}^{\mathbb{P}^{\mathbf{x}}} \biggl[
{ X_1 (T) + \cdots+ X_n
(T) \over X_1 (T) \cdots X_n (T) } \biggr]
\end{equation}
[see Goia (\citeyear{G09}) and Pal (\citeyear{P09}) for a computation of the joint density of
$ X_1 (T),\ldots, X_n (T)$ which leads then
to an explicit computation of $ U(T, \mathbf{x}) $ in
(\ref{E.U}) above, and shows that this function is indeed of class $
\mathcal{C}^{1,2} $].

With $ \zeta=1 $ one computes $ Z(t) =
\prod_{j=1}^n (X_j (0) / X_j (t)) $, therefore
$ \Lambda(t)= ( X(t)/ X (0) )^{n-1} \prod_{j=1}^n (
\mu_j (t)/ \mu_j (0) )
$ as well as\vspace*{1pt}
$ \Lambda_i (t)= ( X(t)/ X (0) )^{n-1}\cdot \break \prod_{j\ne i}  (
\mu_j (t) / \mu_j (0) ) $ for $ i=1,\ldots, n $. Both
representations in (\ref{C.6aa})
hold for the first hitting time of (\ref{Ttau}) in
this case; whereas
$ \mathcal{S}
= \mathcal{T} = \min_{1\le i \le n} \mathcal{T}_i $ as in
(\ref{Ttau})--(\ref{C.6b}), since $ L(t) = ( 1 / Z(t))= ( X(t)
/ X (0) )^{n }  \prod_{j=1}^n ( \mu_j (t)/ \mu_j (0) ) $.

Both (\ref{E.2.a}) and (\ref{E.2.abc}) hold for the example
of (\ref{D.8*a}) with $ h =n-1 $, the first as equality;
from the corollary to Proposition \ref{Proposition2}
and Remark \ref{Remark3}, (\ref{E.3.v}) holds. We recover the result
of Banner and Fernholz (\citeyear{BF08}) on the
existence of arbitrage relative to market of (\ref{D.8*a}) over
\textit{arbitrary} time--horizons.

The diffusion process $
\mathfrak{Y}(\cdot)$ of (\ref{E.4}) takes now the form
%
%
\begin{equation}
\label{D.8**}
d Y_i (t) = Y_i (t)
\,dt + \sqrt{ Y_i (t) \bigl( Y_1
(t) + \cdots+ Y_n (t) \bigr) } \,d W_i (t) .
\end{equation}
The conditions of Bass and Perkins
(\citeyear{BP03}) are satisfied again, though one should compare the ``weak
drift'' $ \widehat{\mathfrak{b}}_i (\mathbf{x}) = x_i \ge0 $
in (\ref{D.8**}), which vanishes for \mbox{$x_i =0$}, with the
``strong drift'' $ \mathfrak{b}_i (\mathbf{x}) =
\kappa(x_1 + \cdots+ x_n) $ for the the original diffusion
$ \mathfrak{X}(\cdot) $ in (\ref{D.8*}), which is strictly
positive on $ [0, \infty)^n\setminus\{ \mathbf{0} \}$.

The corresponding Fichera drifts in (\ref{FicheraToo}), (\ref{Fichera})
are given by $ 2 \widehat{\mathfrak{ f}}_i (\mathbf{ x})= x_i
-( x_1 + \cdots+ x_n )
$, $ 2 \mathfrak{ f}_i (\mathbf{ x}) = \zeta( x_1 +
\cdots+ x_n ) - x_i $,
and $ \mathfrak{ f}_i (\mathbf{ x}) >0> \widehat{\mathfrak{ f}}_i
(\mathbf{ x}) $ hold on $ \{ x_i =0\} \cap\{ \sum_{j \neq i} x_j >0
\} $; from Remark \ref{Remark2} we verify again that the diffusion $ \mathfrak
{X}(\cdot) $ of (\ref{D.8*}) takes values in $ (0,\infty)^n$.

In contrast, the new diffusion $ \mathfrak{Y}(\cdot) $
of (\ref{D.8**}) lives in $ [0, \infty)^n\setminus\{ \mathbf{0}
\} $, and hits the boundary $ \mathcal{O}^n $ of this
nonnegative orthant with positive probability
$ \mathbb{Q}^{\mathbf{x}} [ \mathfrak{T} \le T ] = 1
- U(T,\mathbf{x}) $ for every $ T \in(0, \infty) $. The
positive $ \mathbb{P}^\mathbf{x}$-supermartingale
$ Z(\cdot) X(\cdot) $ is a $ \mathbb{P}^\mathbf{x}$-potential,
for every $ \mathbf{x} \in(0,\infty)^n $. In
this case, the three inequalities of (\ref{AO.b}) hold for every
$ T \in(0, \infty) $: \textit{the local martingales $ Z(\cdot) $,
$ Z(\cdot) X(\cdot) $
and $ Z(\cdot) X_i (\cdot) $, $i=1,\ldots, n $ are all
strict. }

The
model (\ref{D.8*a}) can be cast in the form
(\ref{G.2}) for the relative market weights, as a \textit{multivariate
Jacobi diffusion process} with dynamics
$ d \mu_i (t)= (1 + \zeta) ( 1 - n \mu_i
(t) ) \,dt + \sqrt{ \mu_i (t) } \,d W_i (t)- \mu_i (t)
\sum_{k=1}^n \sqrt{ \mu_k (t) } \,d W_k (t)
$, or
%
%
\begin{equation}
\label{G.9}
d \mu_i (t) = (1 + \zeta) \bigl( 1 -n \mu_i (t) \bigr) \,dt +
\sqrt{ \mu_i (t) \bigl( 1 - \mu_i (t) \bigr) } \,d W^\sharp_i (t)
\end{equation}
with appropriate
Brownian motions $ W^\sharp_1 (\cdot),\ldots,
W^\sharp_n (\cdot) $. Thus, each component $ \mu_i
(\cdot) $ is also a diffusion on the unit interval $
(0,1) $ with local drift $ (1 + \zeta) ( 1 -n y
) $ and local variance $ y (1-y) $ of Wright--Fisher type.
Goia (\citeyear{G09}) studies in detail this multivariate diffusion $
\mu(\cdot) $ based on an extension of the Warren and Yor (\citeyear{WY97}),
Gouri\'eroux and Jasiak (\citeyear{GJ06}) study of skew-products involving Bessel
and Jacobi processes.

From
(\ref{D.8**}), $ Y(\cdot) := Y_1(\cdot) + \cdots+
Y_n(\cdot) $ satisfies the stochastic equation
$ d Y(t) = Y(t) [ dt + d B(t) ] $, where $ B(\cdot)
:= \sum_{j=1}^n \int_0^\cdot\sqrt{Y_j (t) / Y(t) } \,d W_j
(t)$ is Brownian motion; thus $ Y(\cdot) $ a
geometric Brownian motion with drift, under $ \mathbb{Q}^\mathbf{
x} $. The process $ \nu(\cdot) = ( \nu_1
(\cdot),\ldots, \nu_n (\cdot) )' $ of (\ref{G.7}) is related
to the auxiliary diffusion $ \mathfrak{Y} (\cdot) $ of
(\ref{D.8**}) via $ \nu_i (\cdot) = Y_i (\cdot)/ Y(\cdot) $.

The dynamics of these $ \nu_i (\cdot) $'s are easy to
describe in the manner of (\ref{NUs}), namely, $ d \nu_i (t)=
\sqrt{ \nu_i (t) } \,d W_i (t) - \nu_i (t)
\sum_{k=1}^n \sqrt{ \nu_k (t) } \,d W_k (t) $,
or in the notation of (\ref{G.9}):
$ d \nu_i (t) =
\sqrt{ \nu_i (t) ( 1 - \nu_i (t) ) } \,d W^\sharp_i (t) $. Then the
Feller test [e.g., Karatzas and Shreve (\citeyear{KS91}), pages 348--350] ensures
that each $ \nu_i (\cdot) $ hits one of the endpoints of
$(0,1)$ in finite expected time. Thus,
all but one of the $ Y_i (\cdot) $'s eventually get absorbed at zero;
from that time $ \mathfrak{T}_* $ [with
$ \mathbb{E}^{\mathbb{Q}^{\mathbf{x}}}(\mathfrak{T}_*) < \infty$]
onward, the only surviving nonzero component $ Y(\cdot) $
behaves like geometric Brownian motion with drift; in particular, $
\mathfrak{Y} (\cdot) $ never hits
the origin.


\section{Some open questions}\label{sec13}

What conditions, if any,
on the Markovian covariance structure
of Section \ref{sec8} will guarantee that $ \widehat{\pi} (\cdot) $ of
(\ref{F.6}) never
borrows from the money-market, that is, $
\sum_{i=1}^n x_i D_i U(T,\mathbf{x}) \le0 $?
That it is
a portfolio, i.e., that $ \sum_{i=1}^n x_i D_i
U(T,\mathbf{x}) = 0
$ holds? (See Remark \ref{Remark6} for a partial answer.) Or better, that $
\widehat{\pi} (\cdot) $ of (\ref{F.6}) is a long-only portfolio,
meaning that both this condition and $
D_i
( G(T,\mathbf{x}) e^{ H(\mathbf{x})} ) \ge0 $
hold?

Can an iterative method be constructed which converges to the
minimal solution of the parabolic differential inequality
(\ref{E.3}), $ U(0, \cdot) \equiv1 $ and is numerically
implementable [possibly as in Ekstr\"om, Von Sydow and Tysk (\citeyear{EST08})]? How
about a Monte Carlo scheme that computes the quantity $
U(T,\mathbf{x}) $ of (\ref{E.6}) by generating the paths of the
diffusion process $ \mathfrak{Y}(\cdot) $,
then simulating the probability $ \mathbb{Q}^{\mathbf{x}}
[ \mathfrak{T} >T ] $ that
$ \mathfrak{Y}(\cdot) $ does not hit the boundary of the
nonnegative orthant by time $ T $, when started at $
\mathfrak{Y} (0) = \mathbf{x} \in(0, \infty)^n $?

How does $ U (T,\mathbf{x}) $ behave as $ T \rightarrow
\infty$? If it decreases to zero, then at what rate?


\section{Note added in proof}\label{sec14}

In the context of Proposition \ref{Proposition1}, and under the probability measure $
\mathbb{Q} $ of Section \ref{sec71}, the processes $ X_1 (\cdot),\ldots,
X_n (\cdot) $ are real-valued (do not explode) if and only if their
sum $ X(\cdot) $ as in (\ref{2.1}) is real-valued. Now it is fairly
straightforward to check from (\ref{C.9}) that this sum satisfies the equation
\[
d X(t) = X(t) [ d \langle\widetilde{M} \rangle(t) + d
\widetilde{M} (t) ] ,
\]
where the continuous, $ \mathbb{Q}$-local martingale $ \widetilde
{M}(\cdot) $ and its quadratic variation process $ \langle
\widetilde
{M} \rangle(\cdot) $ are given, respectively, as
\[
\widetilde{M}(t) := \sum_{k=1}^n \int_0^t \Biggl( \sum_{i=1}^n \mu
_i (s)
\sigma_{i k} (s) \Biggr) \,d \widetilde{W}_k (s) ,\qquad \langle
\widetilde
{M} \rangle(t)= \int_0^t \mu^{\prime} (s) \alpha(s) \mu(s) \,ds .
\]
Thus by the Dambis--Dubins--Schwarz result [e.g., Karatzas and Shreve
(\citeyear{KS91}), Theorem 3.4.6], for some real-valued $ \mathbb{Q}$-Brownian
motion $ \widetilde{B} (\cdot) $ we have
\[
\log\biggl( { X (t) \over X(0)} \biggr) = \biggl( \widetilde{B}
(u) + { 1 \over2 } u \biggr) \bigg|_{ u = \langle\widetilde{M}
\rangle(t)} ,\qquad 0 \le t < \infty.
\]

It is fairly clear form this representation that a sufficient condition
for the total capitalization process $ X(\cdot) $ to be real-valued,
$ \mathbb{Q}$-a.e., is that this should hold for the quadratic
variation process $ \langle\widetilde{M} \rangle(\cdot) $:
\[
\mathbb{Q} \bigl( \langle\widetilde{M} \rangle(t) <\infty,
\forall t \in[0, \infty) \bigr) = 1 .
\]
In the volatility-stabilized model of Section \ref{sec121} we have $
\alpha
_{ij} (t) = \delta_{ij} / \mu_i (t) $ and thus
$ \langle\widetilde{M} \rangle(t) = \sum_{i=1}^n \int_0^t \mu_i
(s) \,ds =t $, so this condition is clearly satisfied.

\section*{Acknowledgments}
We wish to thank G. \v{Z}itkovi\'c, N. Sesum, E. R. Fernholz, A.
Banner, V. Papathanakos, T. Ichiba and most notably J. Ruf for several
helpful discussions. We are grateful to M. S\^irbu, M. Soner, F.
Delbaen, W. Schachermayer, J. Hugonnier, J. Teichmann and C. Kardaras
for their comments. We are also deeply indebted to the referees and
Associate Editor for their meticulous readings, and for
suggestions that improved this paper greatly.


%
\printaddresses

\end{document}